\documentclass[a4paper,fleqn,usenatbib]{mnras}
\usepackage{graphics,epsfig,txfonts}
\bibpunct{(}{)}{;}{a}{}{,}

\usepackage{graphicx}

\usepackage{longtable}
\usepackage{amssymb}
\usepackage{xcolor}
\usepackage{color}
\usepackage{lipsum}
\usepackage{textcomp}
\usepackage{threeparttable}
\usepackage{enumerate}

\usepackage{placeins}
\usepackage{float}

\usepackage[normalem]{ulem}
\usepackage{soul}
\usepackage{blindtext}
\usepackage[hyphenbreaks]{breakurl}

\newcommand{\oergcm}[1]{$10^{#1}$ erg cm$^{-2}$ s$^{-1}$}
\newcommand{\ergs}[1]{$\times 10^{#1}$ erg s$^{-1}$}
\newcommand{\oergs}[1]{$10^{#1}$ erg s$^{-1}$}
\newcommand{\hcm}[1]{$\times 10^{#1}$ cm$^{-2}$}

\newcommand{\ct}{cts s$^{-1}$}

\newcommand{\ltsima}{$\buildrel < \over \sim$}
\newcommand{\lsim}{\lower.5ex\hbox{\ltsima}}
\newcommand{\gtsima}{$\buildrel > \over \sim$}
\newcommand{\gsim}{\lower.5ex\hbox{\gtsima}}
\newcommand{\msun}{M$_{\odot}$}

\newcommand{\igr}{IGR\,J05007-7047\xspace}
\newcommand{\lxp}{LXP\,38.55\xspace}
\newcommand{\swift}{{\it Swift}\xspace}

\newcommand{\xmm}{{\it XMM-Newton}\xspace}

\newcommand{\cxo}{\hbox{Chandra}\xspace}
\newcommand{\inte}{\hbox{INTEGRAL}\xspace}

\newcommand{\eqb}{\begin{eqnarray}}
\newcommand{\eqe}{\end{eqnarray}}

\begin{document}

\title[Multi-wavelength properties of \hbox{IGR\,J05007-7047} (\lxp)]{Multi-wavelength properties of \hbox{IGR\,J05007-7047} (\lxp) and identification as a Be X-ray binary pulsar in the LMC}
\author[G. Vasilopoulos et al.]{G.~Vasilopoulos$^1$\thanks{Email: gevas@mpe.mpg.de}, F.~Haberl$^1$, C.~Delvaux$^1$, R.~Sturm$^1$, A.~Udalski$^2$\\
$^{1}$Max-Planck-Institut f\"ur extraterrestrische Physik,Giessenbachstra{\ss}e, 85748 Garching, Germany \\
$^2$  Warsaw University Observatory, Al. Ujazdowskie 4, 00-478 Warszawa, Poland}

\date{Received ?? ??? 2016 / Accepted ?? ??? 2016}

\maketitle

\begin{abstract}
We report on the results of a $\sim$40 d multi-wavelength monitoring of the Be X-ray binary system IGR\,J05007-7047 (\lxp).
During that period the system was monitored in the X-rays using the \swift telescope and in the optical with multiple instruments.
When the X-ray luminosity exceeded \oergs{36} we triggered an \xmm ToO observation.
Timing analysis of the photon events collected during the \xmm observation reveals coherent X-ray pulsations with a period of 38.551(3) s (1$\sigma$), making it the 17$^{th}$ known high-mass X-ray binary pulsar in the LMC.
During the outburst, the X-ray spectrum is fitted best with a model composed of an absorbed power law ($\Gamma=0.63$) plus a high-temperature black-body (kT $\sim2$ keV) component.
By analysing $\sim12$~yr of available OGLE optical data we derived a 30.776(5)~d optical period, 
confirming the previously reported X-ray period of the system as its orbital period.
During our X-ray monitoring the system showed limited optical variability while its IR flux varied in phase with the X-ray luminosity, which implies the presence of a disk-like component adding cooler light to the spectral energy distribution of the system. 
\end{abstract}


\maketitle

\section{Introduction}
\label{sec-intro}

Be/X-ray binaries (BeXRBs) are a major subclass of high-mass X-ray binaries (HMXBs) that accord the majority of accreting X-ray pulsar systems. 
These systems are composed of a highly magnetised neutron star (NS), which accretes matter from a non super-giant Be type donor star.
Since a young stellar population is needed in order to form such systems, their number within a galaxy is correlated with the recent star formation. 
BeXRBs are highly variable in the X-rays exhibiting moderate ($L_x\sim$\oergs{36}) or major outbursts ($L_x\ge$\oergs{37}). 
The X-ray variability is attributed to their irregular mass transfer mechanism, as accretion is fuelled by material ejected from the Be star towards its equatorial plane, forming a decretion disk around it. 
The spectral energy distribution (SED) of OBe stars is composed of two distinct
components, a hot component originating from the stellar atmosphere and an 
infra-red (IR) excess produced by the equatorial decretion disk \citep{1994A&A...290..609D}.
For a comprehensive review, of the major characteristics of BeXRBs, we refer the reader to \citet{2011Ap&SS.332....1R}.

The Magellanic Clouds (MCs) offer a unique possibility to study the population of high-energy sources in a whole galaxy. 
Their moderate and well measured distances of $\sim$50~kpc for the Large Magellanic Cloud (LMC) \citep[][]{2013Natur.495...76P} and $\sim$62~kpc for the Small Magellanic Cloud (SMC) \citep[][]{2014ApJ...780...59G}
as well as their low Galactic foreground absorption  ($\sim6\times10^{20}$cm$^{-2}$) makes them ideal targets for studying X-ray binary systems.

Within the last decade, numerous discoveries of new X-ray binaries have been made by monitoring the MCs (and the whole sky) in hard X-rays. 
The IBIS/ISGRI telescope \citep{2003A&A...411L.131U} on board the INTEGRAL satellite \citep{2003A&A...411L...1W} and
the Burst Alert Telescope \citep[BAT,][]{2005SSRv..120..143B} on board the \swift observatory \citep{2004ApJ...611.1005G} have had a significant contribution. 
On the other hand, X-ray surveys of the Magellanic Clouds in the 0.3-10~keV energy band, like the \xmm survey of the LMC (PI: F. Haberl) and the SMC \citep[PI: F. Haberl,][]{2012A&A...545A.128H,2013A&A...558A...3S} or \swift/XRT observations during the UV survey of both MCs \citep[PI: S. Immler,][]{2014AAS...22344230H}, benefit from their better angular resolution for identifying BeXRB candidate systems.
In particular the high sensitivity and sufficient energy resolution of \xmm allows the spectral characterization of the source and the  identification of the optical counterpart.

The discovery of \igr as an X-ray transient was made with \inte in the 17-60 keV band while its identification as a HMXB was based on a \cxo observation \citep{2005A&A...444L..37S}. \citet{2011A&A...529A..30D} reported a 30.77~d X-ray periodic modulation based on \swift/BAT survey data.
The optical counterpart of the X-ray source was identified as USNO-B1.0 0192-0057570 having a spectral type of B2 III \citep{2006A&A...459...21M}.

In the current study we present the results of the X-ray spectral and temporal analysis of an \xmm ToO observation that was triggered during the period of maximum luminosity of a moderate X-ray outburst in November 2014. 
The \xmm observation revealed coherent 38.55~s pulsations. Following the nomenclature introduced by \citet{2005MNRAS.356..502C} for the BeXRB pulsars in the SMC we suggest the alternative name {LXP\,38.55}.
Moreover, we report the results of our optical and X-ray monitoring during an orbital period of the system at the time of the X-ray outburst.  
Finally, we present a complete analysis of the $\sim$12 yr optical light-curve (LC) of the system as observed by the Optical Gravitational Lensing Experiment \citep[OGLE,][]{2015AcA....65....1U}.

\section{Analysis and results of X-ray observations}
\label{sec-observations}

\begin{table*}
\caption{X-ray observations log of \lxp.}
\begin{center}
\scalebox{0.98}{
\begin{threeparttable}
\begin{tabular}{lcccccclc}
\hline\hline  
\#& OBS-ID & Instrument $^a$ &  T\_start & orbital phase $^b$    &exposure $^c$ & count rate $^d$ & UVOT filter $^e$ &  AB magnitudes $^f$   \\
  &        &            &   MJD [d]    &       &[s]       & cts s$^{-1}$    \\
\hline\noalign{\smallskip}    

\xmm       &  &  &  &  & & & & \\
1&  Slew       & EPIC-pn     &  54068.2    &   0.87   & 6.6      & 1.05$\pm$0.3 $^g$ & --&-- \\
2&  0743910301 & EPIC-pn     &  56981.611  &   0.54   &22039    & 1.337$\pm$0.009  & --&-- \\
 &             & EPIC-MOS1   &  56981.611  &   0.54   &23701    & 0.469$\pm$0.006  & --&-- \\
 &             & EPIC-MOS2   &  56981.611  &   0.54   &23672    & 0.442$\pm$0.005  & --&-- \\
 
\noalign{\smallskip}\hline\noalign{\smallskip}
\swift      &  & &   &  & &  & &\\
1 & 00031846004  & XRT   & 56956.025 &   0.71  & 662   &  0.068$\pm$0.014 & u        & 14.46$\pm$0.02   \\
2 & 00031846005  & XRT   & 56966.903 &   0.06 & 1086  &  0.022$\pm$0.006 & uvm2     & 14.46$\pm$0.02   \\
3 & 00031846006  & XRT   & 56968.232 &   0.10  & 869   &  0.019$\pm$0.007 & u        & 14.45$\pm$0.02   \\
4 & 00031846008  & XRT   & 56972.560 &   0.25  & 1251  &  0.033$\pm$0.007 & u        & 14.42$\pm$0.02   \\
5 & 00031846009  & XRT   & 56974.555 &   0.32  & 1036  &  0.038$\pm$0.008 & uvm2     & 14.57$\pm$0.02   \\
6 & 00031846010  & XRT   & 56976.810 &   0.38  & 1076  &  0.087$\pm$0.012 & u        & 14.44$\pm$0.02   \\
7 & 00031846011  & XRT   & 56978.474 &   0.44  & 1061  &  0.140$\pm$0.015 &  uvm2    & 14.44$\pm$0.02   \\
8 & 00031846012  & XRT   & 56980.671 &   0.50  & 966   &  0.130$\pm$0.017 & u        & 14.41$\pm$0.02   \\
9 & 00031846013  & XRT   & 56982.233 &   0.56  & 167   &  0.14 $\pm$0.04  & uvm2     & 14.48$\pm$0.02   \\
10& 00031846014  & XRT   & 56984.545 &   0.64  & 1900  &  0.098$\pm$0.009 & u        & 14.44$\pm$0.02   \\
  &              &       &           &         &       &                  & uvw2     & 14.46$\pm$0.02  \\
11& 00031846015  & XRT   & 56986.661 &   0.70  & 1286  &  0.069$\pm$0.010 & uvm2     & 14.52$\pm$0.02  \\   
12& 00031846017  & XRT   & 56988.077 &   0.75 & 1179  &  0.076$\pm$0.011 & u        & 14.48$\pm$0.02  \\
13& 00031846018  & XRT   & 56990.520 &   0.83 & 867   &  0.032$\pm$0.008 & uvm2     & 14.43$\pm$0.02  \\
14& 00031846019  & XRT   & 56992.583 &   0.90 & 1011  &  0.026$\pm$0.007 & u        & 14.43$\pm$0.02  \\
15& 00031846020  & XRT   & 56994.658 &   0.96 & 4821  &  0.021$\pm$0.009 & uvm2     & 14.44$\pm$0.02  \\ 
16& 00031846021  & XRT   & 56996.444 &   0.02 & 1284  &  0.016$\pm$0.005 & u        & 14.43$\pm$0.02  \\

\noalign{\smallskip}\hline\noalign{\smallskip}
\cxo      &  & &   &  &  & &\\
1&  6271       & ACIS-I     &  53537.125  &  0.61    &   3415      & 0.200$\pm$0.008 $^f$ & --&-- \\

\hline 
\end{tabular}
{
 \tnote{a} Observation setup: For the \xmm ToO, full-frame mode was used for all EPIC instruments, while medium filter was used for the EPIC/MOS cameras and thin filter for the EPIC/PN. For \swift/XRT the photon-counting mode (PC) was used.  \\
 \tnote{b} The zero orbital phase is selected to match Fig. \ref{fig:multi_lc_all} as $\rm MJD - 56965$ d. \\
 \tnote{c} Exposure time for \xmm EPIC cameras, \cxo/ACIS, or total \swift/XRT exposure time of all snapshots taken within the duration of the obs-id. \\
 \tnote{d} Vignetting corrected count rate in the 0.3-10.0 keV energy band.\\
 \tnote{e} UVOT filter used during the \swift/UVOT integration. \\
 \tnote{f} \swift/UVOT AB magnitudes, not corrected for extinction. \\
 \tnote{g} count rate of the slew survey is given tin the 0.2-12 keV band.}
\end{threeparttable}
}
\end{center}
\label{tab:xray-obs}
\end{table*}

\subsection{X-ray observations}   
\label{sec-xobs}

\begin{figure}
  \resizebox{\hsize}{!}{\includegraphics[angle=0,clip=]{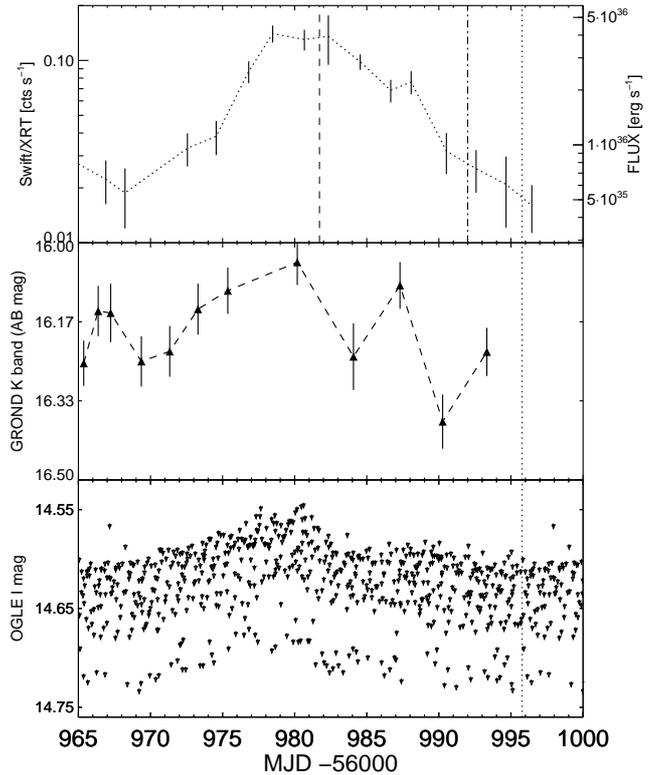}}
\caption{Multi-wavelength light curve of \lxp during a moderate outburst that resulted in a trigger of
an XMM-Newton ToO.
Top panel shows the \swift/XRT light curve during the November 2014 outburst,
middle panel shows the GROND K band light curve during the same period.
Bottom panel shows the total OGLE-IV I band light curve, folded for a period of $\sim30.775~{\rm d}$.
In all the panels the time between MJD~56965 and the vertical dotted line denotes one $\sim30.775~{\rm d}$ optical period.
In the top panel the vertical dashed line indicates the time of the \xmm
ToO. The dash-dotted line indicates the phase at which the \xmm slew survey detection was made.  }
  \label{fig:multi_lc_all}
\end{figure}

We used \swift/XRT to monitor the most promising BeXRB candidate systems of the MCs \citep[for most recent catalogues see:][]{2016A&A...586A..81H,valia15} to search for an X-ray outburst. 
We detected \igr on 2014 October 26, at an X-ray Luminosity of $\sim$2\ergs{36} in the 0.3-10.0 keV band.
We continued monitoring the system with \swift/XRT for a period of 40 days to achieve an X-ray coverage of about one orbital period (30.771 d, see \S \ref{ogle}).
During the maximum X-ray luminosity we triggered one of our granted \xmm anticipated ToO (PI: R.~Sturm) that was performed on 2014 November 20.
Looking at archival data, the system was detected during an \xmm slew \citep{2008A&A...480..611S} on 2006 November 29 (XMMSL1 J050045.3-704441)
at an angular distance of 4.7\arcsec (error 7.7\arcsec) from the \xmm ToO position (see \S \ref{xray_pos}). 
Moreover, the system was also observed with \cxo on 2005 June 16.
The results of the \cxo observation were published in \citet{2005A&A...444L..37S}, but the derived luminosities were based on a power-law model with photon index characteristic of an AGN ($\Gamma=1.4$). 
Thus, we re-analysed the \cxo observation in order to calculate the luminosity of the system at the given time.  
The complete log of the \swift, \xmm and \cxo observations obtained and analysed for the current work is summarised in Table \ref{tab:xray-obs}.
The X-ray, optical and near IR light curves of the system during our monitoring are presented in Fig. \ref{fig:multi_lc_all}.

\subsection{X-ray position}
\label{xray_pos}
The position of \lxp was determined from the \xmm ToO observation that provided the best statistics. X-ray images were created from all the EPIC cameras using the \xmm standard energy sub-bands \citep{2009A&A...493..339W}. Source detection was performed simultaneously on all the images using the SAS task {\tt edetect\_chain}. The source position was determined to R.A. = 05$^{\rm h}$00$^{\rm m}$46\fs06 and Dec. = --70\degr44\arcmin37\farcs4 (J2000), with a $1\sigma$ statistical uncertainty of 0.04\arcsec. The total 1 $\sigma$ positional error, however, is dominated by the remaining systematic uncertainty assumed to be 0.5\arcsec~ \citep[see section 4.3 of][]{2013A&A...558A...3S}.

\subsection{Timing analysis}
\label{sec-time_an}

\begin{figure}
  \resizebox{\hsize}{!}{\includegraphics[angle=0,clip=]{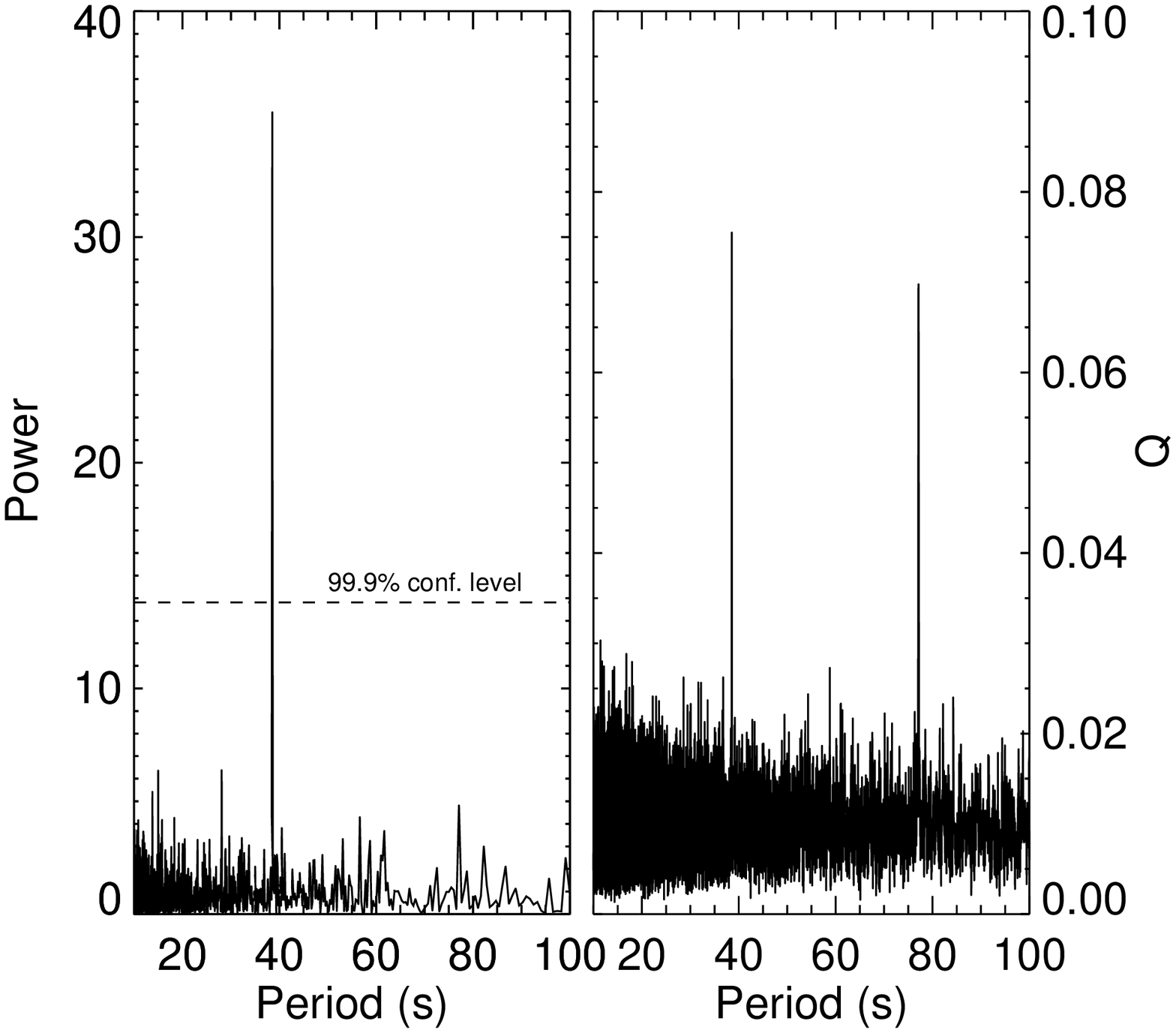}}
    \vspace{-0.5cm}
  \caption{{\it Left:} LS power density spectrum created from the merged EPIC data in the 0.2-10.0 keV energy band. 
           The time binning of the input light curve was 0.1 s. The best-fit period 38.551~s (frequency $\omega\sim0.02594$ Hz) is easily distinguished in the plot. 
           The horizontal line denotes the 99.9\% confidence level of the simulated light-curves.
           {\it Right:} Results of the phase dispersion minimization test (epfold). The epoch-folding test statistics Q is defined as in \citet{1983ApJ...266..160L}. The second peak corresponds to a value two times of the best fit period.} 
  \label{fig:psd}
\end{figure}

We used the SAS task {\tt barycen} to correct the \xmm EPIC event arrival times to the solar-system barycentre. 
To search for a periodic signal we used an epoch folding technique \citep{1990MNRAS.244...93D,1996A&AS..117..197L}.     
We initially searched for a periodic signal in the EPIC-pn event data. The detected signal was longer than the frame time of the EPIC-MOS instrument (2.6 s), thus
in order to increase the signal-to-noise ratio we used the merged event list of EPIC-pn and EPIC-MOS with the common good-time intervals for our timing analysis\footnote{For the analysis the photon arrival times are randomised over the frame time of each instrument.}. 
Additionally, we computed the Lomb-Scargle (LS) periodogram \citep{1982ApJ...263..835S,1986ApJ...302..757H} of the binned X-ray light curve, while performing Monte Carlo white noise simulations to calculate the significance of the derived period. Based on 10000 simulated light curves we conclude that the measured period is significant at a larger than 3$\sigma$ level.
In Fig.~\ref{fig:psd}, we present the inferred power density spectrum of LS method and the results of the epoch folding phase dispersion minimization statistical test. To improve the result of our initial estimation and to estimate the uncertainty of our solution, we followed \citet{2008A&A...489..327H}. 
Based on a Bayesian periodic signal detection method introduced by \citet{1996ApJ...473.1059G}, we determined the pulse period with its 1$\sigma$ uncertainty to  $38.551\pm0.003$ s.

\begin{figure}
  \resizebox{\hsize}{!}{\includegraphics[angle=0,clip=]{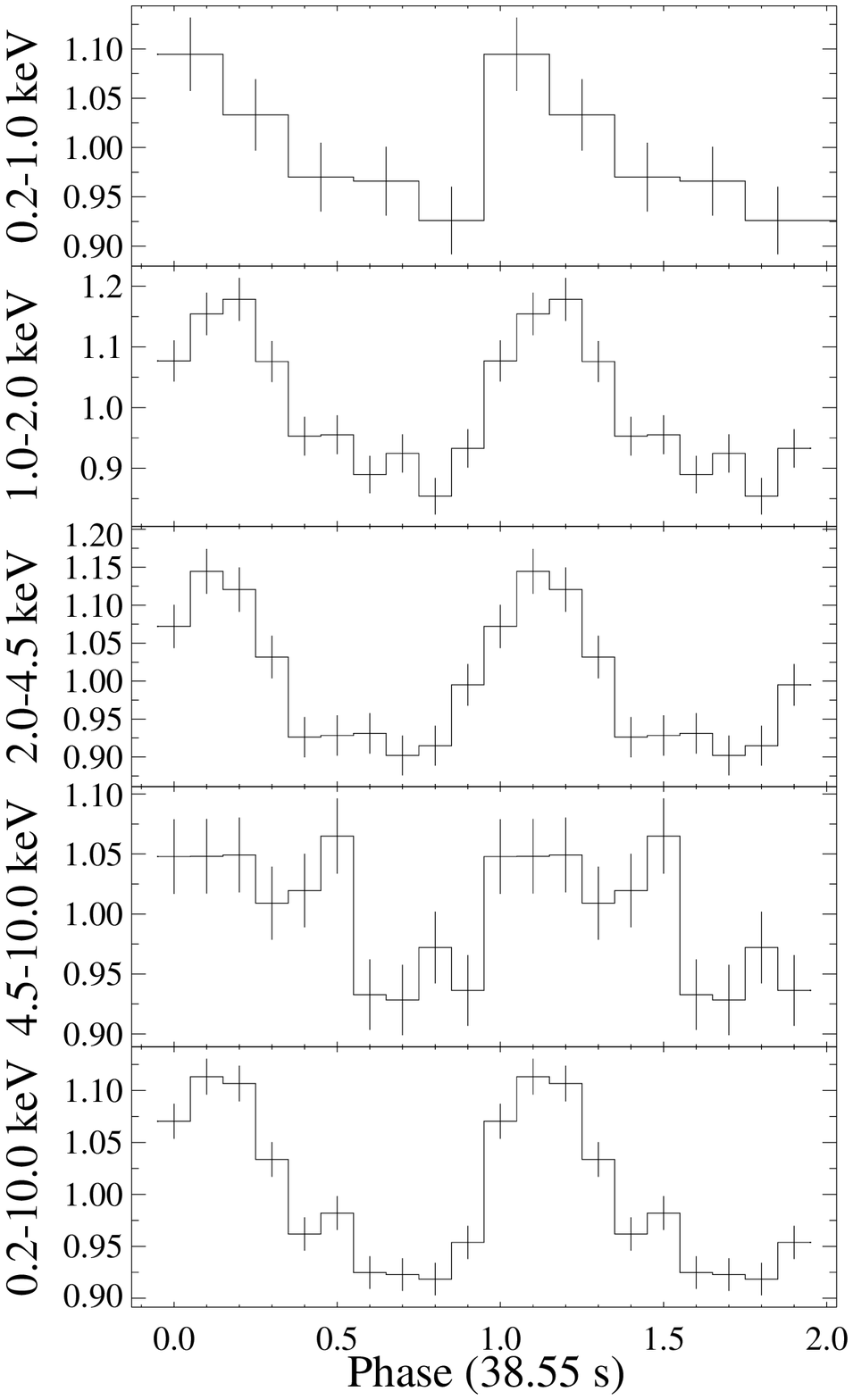}\includegraphics[angle=0,clip=]{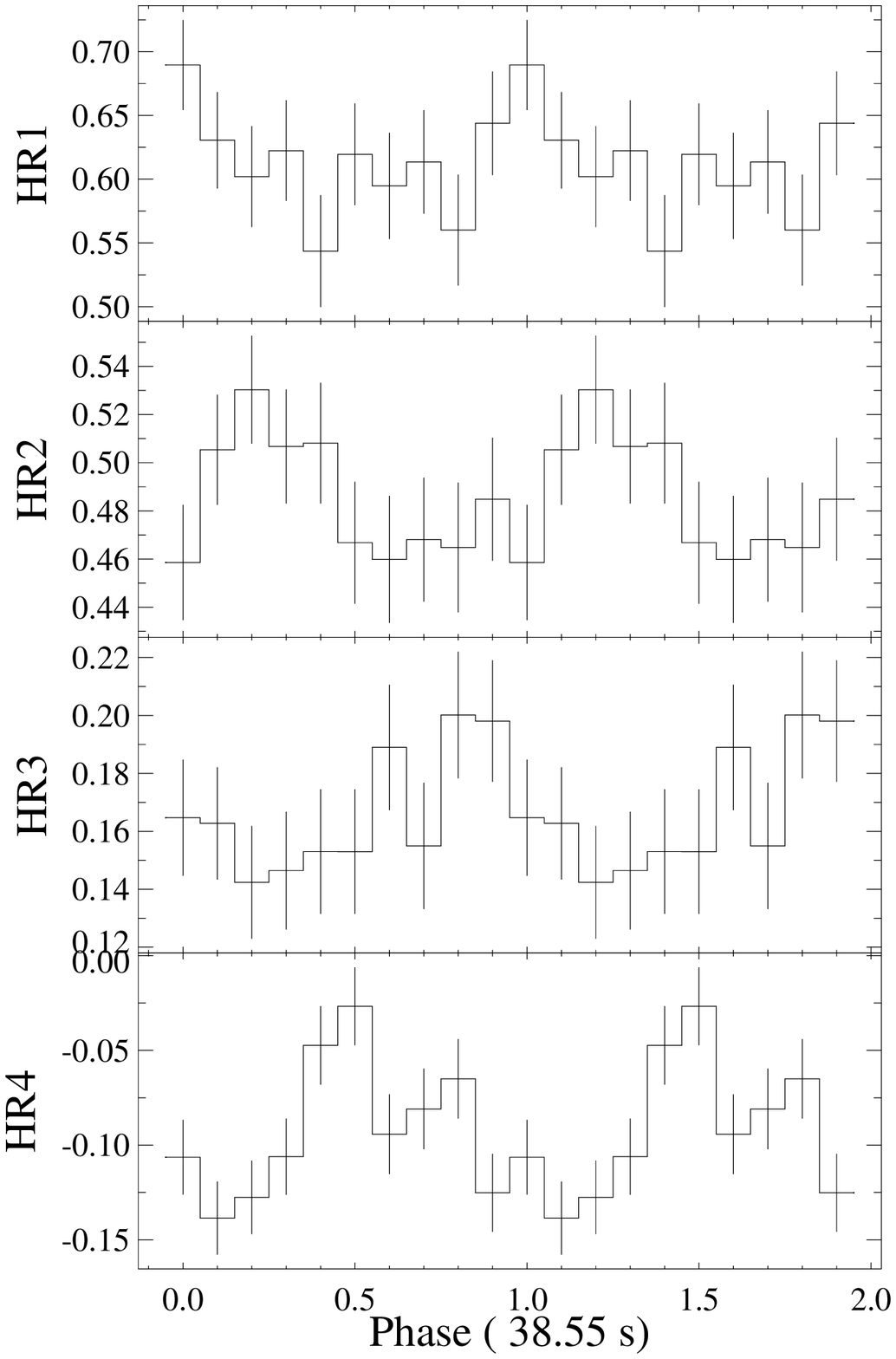}}
 \vspace{-0.5cm}
  \caption{ 
  {\it Left:} Pulse profiles of IGR\,J05007-7047 obtained from the merged EPIC data in different energy bands. 
   (For better statistics the first two standard energy bands were combined in the top panel, while the bottom panel shows all five energy bands combined). 
   The profiles are background-subtracted and normalised to the 
   average count rate (from top to bottom: 0.182, 0.426, 0.597, 0.493 and 1.70 \ct).
  {\it Right:} Hardness ratios derived 
   from the pulse profiles in two neighbouring standard energy bands as a function of pulse phase.
   }    
  \label{fig:pp}
\end{figure}

By using the five standard energy bands (0.2-0.5, 0.5-1.0, 1.0-2.0, 2.0-4.5, 4.5-10 keV), we can define four hardness ratios as
$\rm{HR}_i=(\rm{R}_{\rm{i+1}}-\rm{R}_{\rm{i}})/(\rm{R}_{\rm{i+1}}+\rm{R}_{\rm{i}})$, with R$_{\rm i}$ denoting the background-subtracted count rate in energy band i. The period-folded pulse profiles in the EPIC standard energy bands, together with the hardness ratios derived from the pulse profiles in two adjacent energy bands, are plotted in Fig. \ref{fig:pp}.

\subsection{Spectral analysis}
\label{sec-spec_x}
 \begin{figure}
    \resizebox{\hsize}{!}{\includegraphics[angle=0,clip,trim=0 0 0 0]{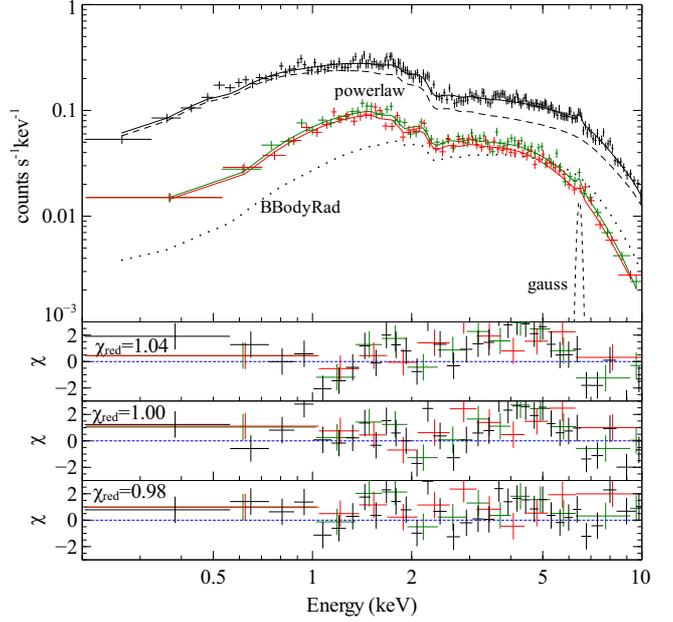}}
     \vspace{-0.5cm}
           \caption{Pulse-phase averaged EPIC spectra of \lxp. The top panel shows the EPIC-pn (black), EPIC-MOS1 (red) and EPIC-MOS2 (green) spectra, together with a best-fit model (solid lines) composed of an absorbed power law with photon index $\Gamma=$0.63, a black body with 2.04 keV temperature(dotted line), and a Gaussian line at 6.46 keV.
           The second panel shows the residuals of an absorbed power law model ($\Gamma=$0.58).
           The third panel shows the residuals for the fitted model of an absorbed power law ($\Gamma=$0.64) plus multi-temperature disk black body model ( kT=0.1 keV ). 
           Bottom panel shows the residuals of the best fit model plotted in the top panel.
           The residuals were rebinned for clarity while the spectra in the top
	   panel show the binning used for the spectral fit.
           We note that residuals at $\sim$2.2~keV are an instrumental feature, related to the Au layer on the mirrors.}
   \label{fig:spec1}
 \end{figure}

We used {\tt xspec} \citep[version 12.8.2, ][]{1996ASPC..101...17A} to perform the X-ray spectral analysis. The \xmm/EPIC cameras provided a total of $\sim$ 35k counts enabling us to use $\chi^2$ statistics in the fitting procedure. The three spectra were fitted simultaneously with the same model with the addition of a scaling factor to account for instrumental differences.
For the EPIC-pn we fixed the scaling factor at 1, while for both EPIC-MOS, we obtained values of $C_{MOS1}=1.03\pm0.02$ and $C_{MOS2}=0.97\pm0.02$, which is consistent with the expected values \citep[see ][or the latest version of the \xmm calibration manual\footnote{\url{http://xmm2.esac.esa.int/external/xmm\_sw\_cal/calib/cross\_cal/}}]{2006ESASP.604..937S}.
The X-ray absorption was modelled using the {\tt tbnew} code, a new and improved version of the X-ray absorption model {\tt tbabs} \citep{2000ApJ...542..914W}, while the Atomic Cross Sections were adopted from \citet{1996ApJ...465..487V}.
The photo-electric absorption was modelled as a combination of Galactic foreground absorption and an additional column density accounting for both the interstellar medium of the LMC and the intrinsic absorption of the source.
For the Galactic photo-electric absorption we used a fixed column density of N$_{\rm H{\rm, GAL}}$ = 0.847\hcm{21} \citep{1990ARA&A..28..215D}, with abundances according to \citet{2000ApJ...542..914W}.
We note that N$_{\rm H{\rm, GAL}}$ was not fixed to the value provided by the Leiden/Argentine/Bonn (LAB) Survey of Galactic HI \citep[1.35\hcm{21},][]{2005A&A...440..775K}, since this value corresponds to the total Galactic and LMC column density due to the filter used in the survey. 
The LMC intrinsic column density $\rm N_{\rm{H},\rm{LMC}}$ was left as a free parameter with abundances of 0.49  for elements heavier than helium \citep{2002A&A...396...53R}.
All the uncertainties were calculated based on a  $\Delta\chi^2$ statistic of 2.706, equivalent to a 90\% confidence level for one parameter of interest.

The spectra are well fitted by an absorbed power law (photon index $\Gamma\sim$0.58, $x^2_{\rm red}=1.05$), however the addition of an additional component can improve the fit at low energies and smooth out the residuals at higher energies (Fig. \ref{fig:spec1}). 
It is not uncommon for HMXBs to show an additional component in their X-ray spectrum. \citet{2004ApJ...614..881H} provided several interpretations for the origin of such emission that they refer to as soft excess. In the case of intermediate X-ray luminosity systems like \lxp, this excess can be a result of reprocessing of hard X-rays from the NS by optically thick accreting material located at the inner edge of the accretion disk, by photo-ionized or collisionally heated diffuse gas, or thermal emission from the NS surface. Moreover, \citet{2013arXiv1301.5120L} showed that many X-ray pulsars show a spectral feature that can be interpreted as black body emission from a small region, most likely the polar cap of the NS. 
Multiple components may contribute to the spectrum of \lxp. 
Although the simple power-law fit is already acceptable and the statistical quality of the EPIC spectra does not require additional model components, 
we test if emission components which are expected to exist, improve the spectral fit. 
We restrict the model to a maximum of  two continuum components, i.e. a combination of the power law with a thermal component. 
We found two solutions which improve the fit at a) low energies (mainly below $\sim$1.5 keV) and b) higher energies (above $\sim$2 keV).

For case a) both a single-temperature black-body (e.g. {\tt BBodyRad} in {\tt xspec}) or a multi-temperature accretion disk (e.g. {\tt EzdiskBB} in {\tt xspec}) with temperature ($\sim0.1$~keV) provide the same improvement in fit quality. 
It is however important to note that each of the models implies a different physical origin for the radiation process. 
As the normalization parameter of both components translates to an emission region with radius\footnote{$R=\sqrt{Norm/\cos{i}}*Dist*f^2$, where $R$ is the inner radius of the disk in km, $Dist$ is the distance to the source in units of 10 kpc, $i$ is the inclination, and $f=1.4$ is the color to effective temperature ratio} of $\sim$200 km, it is more physical to attribute the emission originating from a region close to the inner radius of an accretion disk rather than a hot accretion column that reaches the NS surface. 
The statistical improvement of the fit compared to an absorbed power-low model was estimated with the {\tt ftest} (probability for a value drawn from the F-distribution) and was found to be $\sim$8~$\sigma$.

The best fit is provided by the combination of an absorbed power-low plus a high-temperature black-body component with temperature $\sim$2.04 keV (case b). This model improves significantly the residuals of the fit at higher energies and provides a statistical improvement above a 3.5~$\sigma$ level compared the low temperature black-body plus power law model.
There is clear evidence of an emission line present at 6.4 keV that originates from neutral Fe (K$_\alpha$ line). The width of the line ($\sigma\sim10^{-5}$~keV) was found to be bounded by the spectral resolution of the instrument, thus we fixed the width of the line to zero since an exact value could not be resolved. 
The parameters for the models described above are given in Table \ref{tab:spectra}.

We note that both the low and high temperature black-body components improve the residuals on different parts of the X-ray spectrum and account for different physical processes, thus in principle they could co-exist in the system.
However, modelling the spectrum with a combination of  three continuum
components is not statistically justified. Moreover, the use of three continuum components together with the column density as free parameter adds degeneracy to the solution. While we cannot exclude the existence of more components in the spectrum, the available statistics does not justify the use of a more complicated spectral model.

\begin{figure}
  \resizebox{\hsize}{!}{\includegraphics[angle=0,clip=]{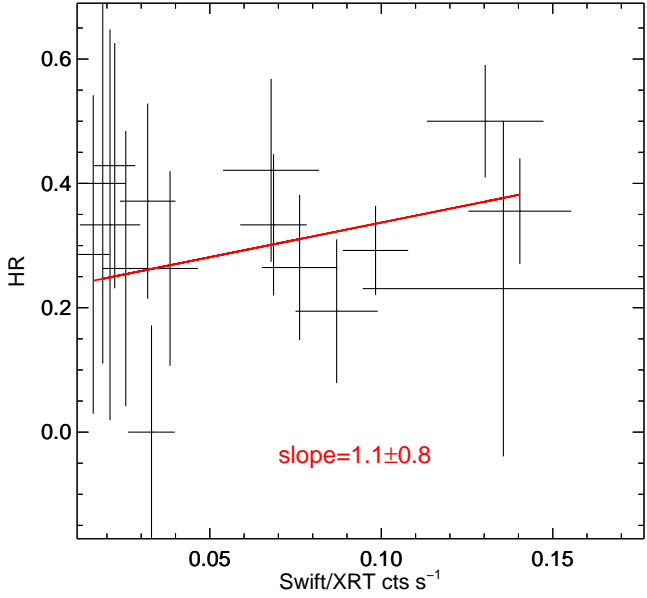}}
   \vspace{-0.5cm}
  \caption{X-ray HR vs \swift/XRT count rate (0.3-10 keV). HR was computed based on the total counts in the low (0.3-2.0 keV) and high (2.0-10 keV) energy band for the duration of the monitoring.
  HR is defined again as the ratio of the differences over the sum of the two energy bands, i.e.  $\rm{HR}=(\rm{R}_{\rm{h}}-\rm{R}_{\rm{l}})/(\rm{R}_{\rm{h}}+\rm{R}_{\rm{l}})$.}
  \label{fig:swift_HR}
\end{figure}

The \swift/XRT observations could not provide sufficient statistics for modelling the X-ray spectrum, we were however able to derive values for the hardness of the spectrum in two adjacent energy bands (0.3-2.0 keV vs. 2.0-10 keV) and compare this quantity along the evolution of the burst. 
There was no significant evidence for a monotonous change in the hardness of the spectrum with luminosity along the $\sim$40 day monitoring cycle (see Fig.\ref{fig:swift_HR}), but a rather weak trend of the system becoming harder with increasing luminosity. 

\begin{table}
\caption{results of the X-ray spectral modelling}
\begin{center}
  \scalebox{0.9}{
 \begin{threeparttable}
\begin{tabular}{lccc}
\hline\hline
Component & Parameter & Value & units \\
\hline\noalign{\smallskip}
\multicolumn{4}{l}{Power-law plus High Energy Black body (T$>$1 keV)}\\
\multicolumn{4}{l}{tbnew*tbnew*(gaussian + bbodyrad + powerlaw)}\\
\hline\noalign{\smallskip}
tbnew    &  nH      &  0.033$^{+0.021}_{-0.018}$                  &  $10^{22} cm^{-2}$      \\ \noalign{\smallskip}    
gaussian &  LineE   &  6.46$^{+0.04}_{-0.04}$          &    keV     \\ \noalign{\smallskip}        
         &  EqWidth   &  0.047$^{+0.04}_{-0.10}$       &    keV     \\ \noalign{\smallskip}                
bbodyrad &  $kT$ &  2.04$^{+0.28}_{-0.3}$ &     keV    \\ \noalign{\smallskip}                 
         &  norm    &  0.020$\pm$0.005          &            \\ \noalign{\smallskip}    
         &  R    &  0.7$^{+0.08}_{-0.09}$          &  km          \\ \noalign{\smallskip}  
powerlaw &  PhoIndex&  0.63$^{+0.12}_{-0.08}$                 &            \\              \noalign{\smallskip}    
         &  Flux/Norm$^{a}$    &  (7.7$^{+1.0}_{-1.3}$)     &       \oergcm{-12}     \\              \noalign{\smallskip}    
\hline
\multicolumn{2}{l}{Observed Flux$^{b}$}          &     1.03$\times 10^{-11}$             &  erg cm$^{-2}$ s$^{-1}$          \\
\multicolumn{2}{l}{Unabsorbed Luminosity$^{c}$}  &    $3.03\pm0.08\times 10^{36}$         &  erg s$^{-1}$          \\ %
\hline
         & $\chi^2$/DOF &     0.98/1174               &            \\
\hline\hline
\multicolumn{4}{l}{Power-law plus Low Energy Black body (T$<$1 keV)}\\
\multicolumn{4}{l}{tbnew*tbnew*(gaussian + bbodyrad + powerlaw)}\\
\hline\noalign{\smallskip}
tbnew    &  nH      &  0.31$\pm0.06$                  &  $10^{22} cm^{-2}$      \\ \noalign{\smallskip}    
gaussian &  LineE   &  6.46$^{+0.04}_{-0.04}$          &    keV     \\ \noalign{\smallskip}        
         &  EqWidth   &  0.047$^{+0.04}_{-0.10}$       &    keV     \\ \noalign{\smallskip}                 
bbodyrad &  $kT$ &  0.088$^{+0.007}_{-0.006}$ &     keV    \\ \noalign{\smallskip}                 
         &  norm    &  2300$^{+2500}_{-1300}$          &            \\ \noalign{\smallskip}     
         &  R    &  240$^{+100}_{-80}$         &  km          \\ \noalign{\smallskip}  
powerlaw &  PhoIndex&  0.60$\pm0.03$                 &            \\              \noalign{\smallskip}    
         &  Flux/Norm$^{a}$    &  (10.92$\pm$0.15)             &     \oergcm{-12}       \\              \noalign{\smallskip}    
\hline
\multicolumn{2}{l}{Observed Flux$^{b}$}          &     1.05$\times 10^{-11}$             &  erg cm$^{-2}$ s$^{-1}$          \\
\multicolumn{2}{l}{Unabsorbed Luminosity$^{c}$}  &    $3.38^{+0.11}_{-0.13}\times 10^{36}$         &  erg s$^{-1}$          \\
\hline
         & $\chi^2$/DOF &     1.00/1174               &            \\
\hline\hline
\multicolumn{4}{l}{Power-law plus low Energy multi-temperature disk Black body (T$<$1 keV)}\\
\multicolumn{4}{l}{tbnew*tbnew*(gaussian + EzdiskBB + powerlaw)}\\
\hline\noalign{\smallskip}
tbnew    &  nH      &  0.34$^{+0.05}_{-0.08}$                   &  $10^{22} cm^{-2}$      \\ \noalign{\smallskip}    
gaussian &  LineE   &  6.46$^{+0.04}_{-0.04}$        &    keV     \\ \noalign{\smallskip}        
         &  EqWidth   &  0.047$^{+0.04}_{-0.10}$                      &    keV     \\ \noalign{\smallskip}                  
EzdiskBB &  $kT_{\rm max}$ &  0.099$^{+0.011}_{-0.006}$ &     keV    \\ \noalign{\smallskip}                 
         &  norm    &  340$^{+300}_{-210}$          &            \\ \noalign{\smallskip} 
         &  inner R    &  215$\mp80$          &  km          \\ \noalign{\smallskip}  
powerlaw &  PhoIndex&  0.605$\pm0.03$                 &            \\              \noalign{\smallskip}    
         &  Flux/Norm$^{a}$   &  (10.92$\pm$0.16)              &  \oergcm{-12}          \\              \noalign{\smallskip}    
\hline
\multicolumn{2}{l}{Observed Flux$^{b}$}          &     1.05$\times 10^{-11}$             &  erg cm$^{-2}$ s$^{-1}$          \\
\multicolumn{2}{l}{Unabsorbed Luminosity$^{c}$}  &    $3.48^{+0.12}_{-0.13}\times 10^{36}$         &  erg s$^{-1}$          \\
\hline
         & $\chi^2$/DOF &     1.00/1174               &            \\
 \hline
\end{tabular}
\tnote{a} Unabsorbed flux of the power-law component in the 0.3-10. keV energy band. \\
\tnote{b} Absorbed flux of the fitted model in the 0.3-10. keV energy band. \\
\tnote{c} Unabsorbed Luminosity of the fitted model in the 0.3-10. keV energy band, for a distance of 50 kpc \citep{2013Natur.495...76P}. \\
\end{threeparttable}
 }
\end{center}
\label{tab:spectra}
\end{table}


\section{Analysis and results of optical data}
\label{sec:datareduction_optical}

\subsection{SWIFT UVOT}

Our \swift monitoring enabled us to measure the UV magnitudes (using the \swift UV/Optical Telescope, UVOT). 
We used the default filter of the day for the \swift/UVOT instrument setup.
This resulted in nine observations performed with the U filter (central wavelength 3465 \AA), seven with the UVM2 (2246 \AA)  and one with the UVW2 (1928 \AA).
To derive the systems UV magnitudes we used the {\tt uvotsource} tool. A 5\arcsec\ radius was used for performing aperture  photometry for all filters.
No significant variation was measured in any of the filters; the mean values for the derived magnitudes were: 14.23$\pm$0.02 for the U,
14.12$\pm$0.05 for the UVM2 and 14.166$\pm$0.06 for the UVW2 filter.

\subsection{GROND}
\lxp was observed thirteen times between 2014 November 4 and December 2 with the Gamma-Ray Burst Optical/Near-Infrared Detector 
\citep[GROND;][]{2008PASP..120..405G} at the MPI/ESO 2.2~m telescope (La Silla, Chile). Seven bands 
were used, providing a coverage in both optical (g', r', i' and z') and near-IR (J, H and K$_{s}$) wavelengths. 
Each individual observation consists of 24 dithered exposures of 10~s in the near-IR and 4 dithered exposures in the optical with various 
exposure times (35~s in the eight first epochs and 66~s in the five last epochs), taken at a mean airmass of 1.3  and a mean seeing of 1.4".

\begin{figure}
 \resizebox{\hsize}{!}{\includegraphics[angle=0,clip=]{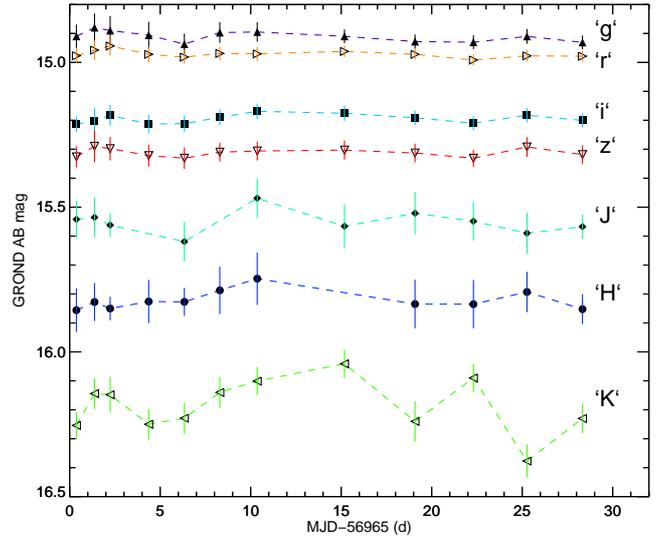}}
     \vspace{-0.5cm}
  \caption{GROND light curves in seven filters.}
  \label{fig:grond_lc_all}
\end{figure}

Single dithered exposures were
reduced (bias subtraction, flat-fielding, distortion correction) and stacked using standard
IRAF\footnote{IRAF is distributed by the National Optical Astronomy Observatories, which are operated by the Association of Universities 
for Research in Astronomy, Inc., under cooperative agreement with the National
Science Foundation.} tasks implemented in the GROND pipeline \citep{2008ApJ...685..376K,2008AIPC.1000..227Y}.
The astrometry calibration was computed on single exposures against stars selected from the
USNO-B1.0 catalogue \citep{2003AJ....125..984M} in the optical bands and the 2MASS catalogue
\citep{2006AJ....131.1163S} in the near-IR bands, yielding an accuracy of 0.3" with respect to the
chosen reference frame. The photometric calibration in the optical was computed against a
close-by field from the Sloan Digital Sky Survey \citep{2000AJ....120.1579Y} at $\delta = -10\deg$. This was observed five 
times over the entire observation run (including the first night) under photometric conditions. From the calibrated images we extracted a grid
of secondary photometric calibrators for direct on-the-frame calibration on the subsequent
nights. In the near-IR, the photometric calibration was computed against 2MASS stars
identified in the GROND field of view. The accuracy of the absolute photometry calibration
was 0.03 mag in g', r', i' and z', 0.07 mag in J and H, and 0.05 mag in K$_{s}$.
The seven band light curves derived from the GROND observation is presented in Fig. \ref{fig:grond_lc_all}.

\begin{figure}
  \resizebox{\hsize}{!}{\includegraphics[angle=0,clip=]{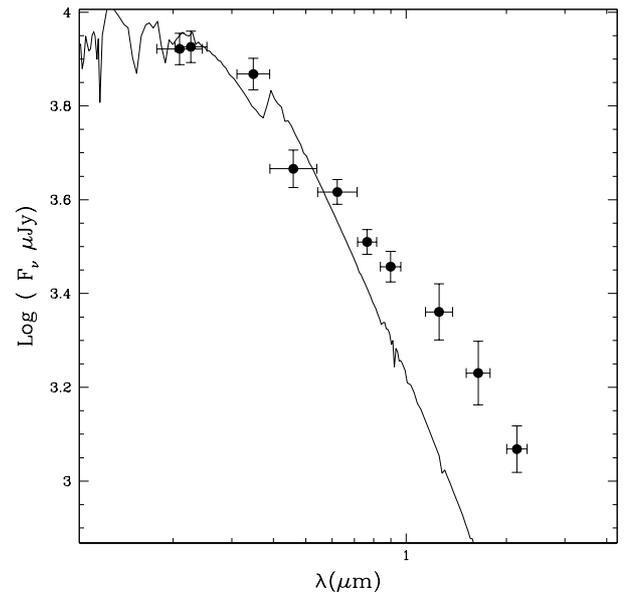}}
  \vspace{-0.5cm}
  \caption{Spectral energy distribution of \lxp. From left to right the three \swift/UVOT and the seven GROND optical and NIR fluxes are plotted. The plotted line corresponds to the Kurucz model with T=22000 K, $\log{z}=-0.05$ and $\log{g}=3.5$.}
  \label{fig:grond_SED}
\end{figure}

Having simultaneously obtained the photometry of the optical companion from
NIR to optical, we can construct its SED and look for an IR excess. 
As BeXRBs are intrinsically variable in the optical, having both orbital and super-orbital variability, GROND data offer us the unique opportunity to study the SED of such systems.  
Moreover, since our \swift/UVOT fluxes showed no significant variability (compared to the NIR fluctuations) we can supplement the GROND magnitudes with the UV data from \swift, constructing an SED with  (quasi-)simultaneous observations.
To de-redden the optical observations, we used the extinction curve given by \citet{1999ApJ...515..128M}, which is most appropriate for the LMC. 
The foreground Galactic reddening was set to $E(B-V)_{\mathrm{Gal}} = 0.1062$\,mag \citep{2011ApJ...737..103S}.
We used the Kurucz models of stellar atmospheres \citep{1979ApJS...40....1K,1997A&A...318..841C} to fit the observed photometry, while constraining the metallicity to that of the LMC ([Fe/H]=-0.5).

\begin{table}
\caption{NIR excess attributed to the Be disk.}
\begin{center}
\scalebox{0.9}{
\begin{threeparttable}
\begin{tabular}{cccc}
\hline\hline
filter &  central wavelength & $F_{\rm observed}/F_{\rm template}$ $^a$ & $F_{\rm excess}$ \\ \noalign{\smallskip}
  & \AA &    & ${\mu}Jy$   \\  \noalign{\smallskip}
\hline\noalign{\smallskip}
i & 7640.0   & $1.24^{+0.08}_{-0.07}$ & $582^{+180}_{-170}$ \\  \noalign{\smallskip}
z & 8989.3   & $1.42^{+0.10}_{-0.10}$ & $779^{+190}_{-180}$ \\  \noalign{\smallskip}
G & 12560.9  & $1.88^{+0.25}_{-0.22}$ & $988^{+290}_{-250}$ \\  \noalign{\smallskip}
H & 16467.1  & $2.14^{+0.32}_{-0.28}$ & $848^{+240}_{-210}$ \\  \noalign{\smallskip}
K & 21512.4  & $2.20^{+0.24}_{-0.21}$ & $625^{+120}_{-110}$ \\  \noalign{\smallskip}
\hline
\end{tabular}
\tnote{a} The ratio was computed based on the best fit model and the observed flux during the first night. The uncertainties were computed accounting for both the observational uncertainties and the systems variability.  \\
\end{threeparttable}
}
\end{center}
\label{tab:sed}
\end{table}

The spectral types of the optical counterparts of BeXRBs in the SMC are reported to be in the range between O5 and down to B9 \citep{2005MNRAS.356..502C}.
However, in the new compilation of spectral types by \citet{2016A&A...586A..81H} only stars with uncertain classification may have types as late as B5. There are no B6, B7 or B8 counterparts and only one B9 which may cast doubts on the spectral classification or the correct identification. This makes the spectral type distribution more similar to that of the Galaxy which seems to end at B3.
A B2 III star, like our system in study, is expected to have a temperature of $\sim$21500 K \citep{2000asqu.book.....C}, thus we limit the temperature parameter to the range of 20000-24000~K. 
We note that the decretion disk of a Be star can account up to 50\% of its total luminosity \citep{2006A&A...456.1027D}, thus stellar atmosphere models are expected to generally produce poor fits to the variable SED of those systems. 
A brute force fit of a stellar atmosphere model to our data without accounting for any prior knowledge of the systems properties (spectral type or distance) produces questionable results.
This method delivers a best fit model of a much cooler than expected star (T$\sim$10000 K) and at a much closer distance than that of the LMC.
By limiting the temperature parameter  search grid to the range of 20000-24000~K, and by excluding the optical and NIR GROND data from the fit we can achieve a good fit, where the normalization and the radius of the star describe our prior assumptions. 
The best derived model was that of a star with temperature of 22000 K and surface gravity $\log{g}=3.5$.
The fitted SED of the system is presented in Fig. \ref{fig:grond_SED}.
From the fitted model it is clear that the contribution of the disk is not restricted to the NIR magnitudes but affects the optical emission of the system as well.

To stress the importance of UV measurements to the SED fitting we repeated the above exercise by using only the GROND magnitudes.
We then get a best fit model of a star with a temperature range of 8500-9500 K, with a normalization pointing to a source located much closer than the LMC distance. 
This suggests the presence of a cool disk that contributes to the SED of the B star.
Further investigation of the contribution of the disk in the SED of the star lies beyond the scope of the present work as it requires precise modelling of the decretion disk.

\subsection{OGLE}
 \label{ogle}
 
The Optical Gravitational Lensing Experiment (OGLE) started its initial observations in 1992 \citep{1992AcA....42..253U} and continues observing till today \citep[OGLE IV,][]{2015AcA....65....1U}. Observations were made with the 1.3~m Warsaw telescope at Las Campanas Observatory, Chile. Images are taken in the V and I filter pass-bands, while the data reduction is described in \citet{2015AcA....65....1U}. 
OGLE photometric magnitudes are calibrated to the standard VI system.

\begin{figure*}
 \resizebox{\hsize}{!}{\includegraphics[angle=0,clip=]{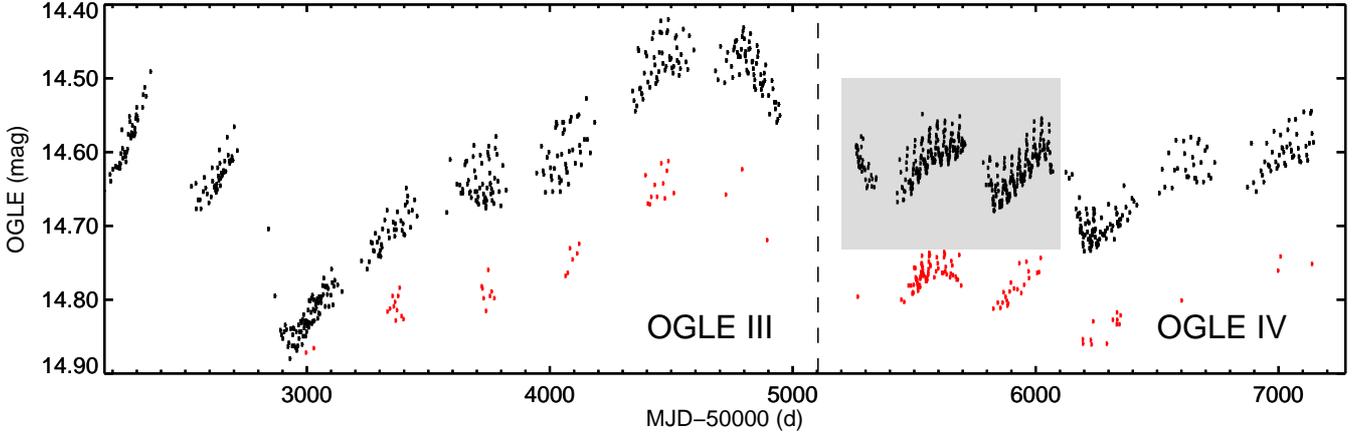}}
    \vspace{-0.5cm}
  \caption{OGLE I (black) \& V (red) band light curves. Data obtained during the OGLE-III and IV phases are separated by the vertical dashed line. The grey shaded area marks the first 900 d of the OGLE-IV phase when the system was observed more frequently. The $\sim$30.77 d period is clearly seen within this interval.}
  \label{fig:ogle_lc1}
\end{figure*}

\begin{figure}
 \resizebox{\hsize}{!}{\includegraphics[angle=0,clip=]{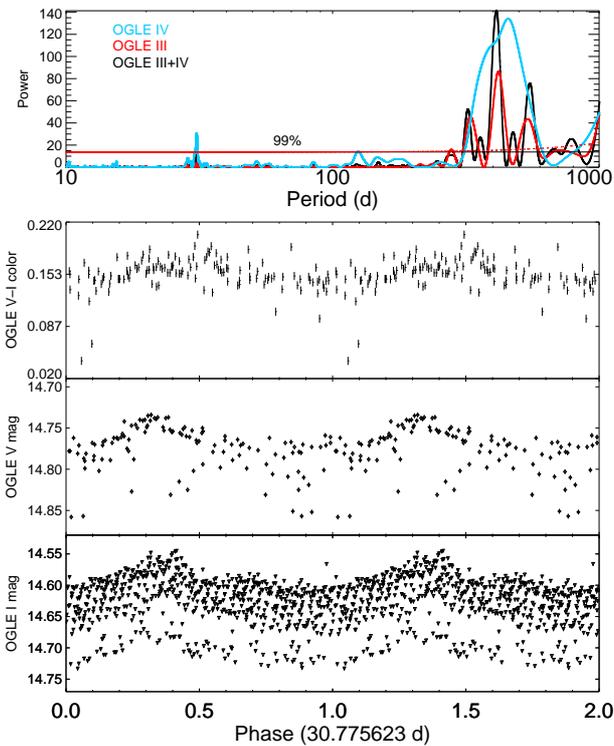}}
   \vspace{-0.5cm}
  \caption{
  Upper panel: Lomb-Scargle periodorgam of the OGLE III and IV I band photometric data of \lxp. Horizontal lines indicate the 99\% confidence level for significant periods.
  Bottom three pannels: OGLE V-I, V \& I band phase folded light curves.
  In order to avoid confusion only data from OGLE-IV epoch were used.
  Throughout the current study all folded light curves are produced with a the same folded period as in this plot ($\sim$30.776 d) and zero phase of 56965 d.
  }
  \label{fig:ogle_lc2}
\end{figure}

In the present work, we used the OGLE I and V band magnitudes for the counterpart of \lxp that were taken between Modified Julian dates (MJD) 52166 and 57275. The latest I and V band light curves are shown in Fig. \ref{fig:ogle_lc1}. The photometric data can be downloaded from the OGLE-IV real time monitoring of X-ray sources web-page\footnote{{\tt XROM}: \url{http://ogle.astrouw.edu.pl/ogle4/xrom/xrom.html}} \citep{2008AcA....58..187U}. 
For the OGLE I band magnitudes we computed periodograms by using the Lomb-Scargle algorithm (see Fig. \ref{fig:ogle_lc2}). 
We searched for periodicities up to half of the total OGLE observing period, which at the time of the analysis was $\sim$5108 d.

By using the $\sim$14 year of available OGLE observations in the I band we were able to derive a refined orbital period of 30.776$\pm$0.005~d (see Fig. \ref{fig:ogle_lc2}).
Additionally, a $410\pm18~\rm d$ super-orbital quasi-periodic modulation can be seen in the data.
We note that the super-orbital modulation should not be treated as a periodic signal like the orbital modulation, its nature is probably related to the precession of the Be disk.
Period uncertainties were computed using the bootstrap method and repeating the calculations for 10000 light curves.
Both orbital and super-orbital modulations are confirmed well above a 99\% confidence level by performing 10000 white and and red noise \citep[using REDFIT,][]{2002CG.....28..421S}\footnote{REDFIT: \url{http://www.ncdc.noaa.gov/paleo/softlib/redfit/redfit.html}} simulations.
Due to the limited OGLE coverage ($\sim12$ super-orbital periods) and the gaps in the light curve the uncertainty on the super-orbital solution is high.
Following equation 14 of \citet{1986ApJ...302..757H} we estimate an uncertainty of about $\sim40$ d for the super-orbital solution.
This value is different from the uncertainty computed by the bootstrap method and is only used as an estimator of the effect of the length of the data to the determination of the periodic solution. 
Observations spanning over a N times larger interval would decrease this uncertainty by a factor of $\sim{N^{3/2}}$, e.g. by doubling the coverage we would only improve the uncertainty to ~14 d.
The shape of the $\sim410~\rm d$ fundamental peak of the Lomb-Scargle periodogram is affected by  both the length of the data set \citep[see Fig. 6 \& 7 of][]{2011MNRAS.413.1600R} and the $365~\rm d$ yearly periodicity that is expected to be found in the OGLE data. 
The final super-orbital period was derived by performing Monte Carlo simulations and fitting the fundamental peak of the Lomb-Scargle periodogram with a Gaussian profile.   
During the OGLE-III phase (MJD $<$ 55000 d) the system exhibits high long term variability in addition to its orbital and super-orbital periodicities.
By analysing data from multiple epochs (1-2 y intervals) we conclude that there has been no statistically significant change in the orbital period within the investigated time intervals.
We note that the orbital period is detected during the OGLE-III phase only at a low significance level ($\sim$50\% confidence level) and it can be easily dismissed by automatic detection algorithms.

\begin{figure}
  \resizebox{\hsize}{!}{\includegraphics[angle=0,clip=]{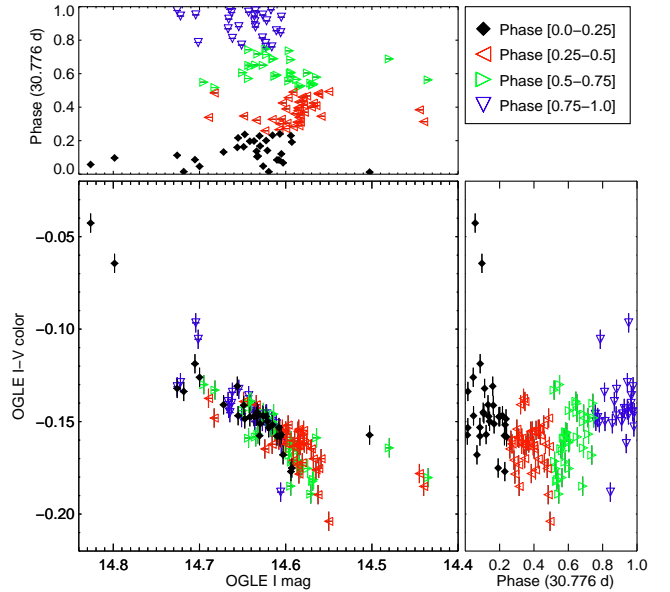}}
    \vspace{-0.5cm}
  \caption{\lxp OGLE colour magnitude diagram.
  For comparison the phase folded I-V colour light curve is plotted on the right, while the phase folded I band light curve is plotted on top of the plot.   
  Both light curves are folded for the $\sim$30.776 d period.
    The group of observations performed between ``MJD 56100`` and ``MJD 56500`` when the optical luminosity of the system was at its lowest stage follows the same modulation in lower luminosities, see group of points with I mag $>$ 14.68 on the left of the main plot.
   }
  \label{fig:ogle_lc_all}
\end{figure}

Figure \ref{fig:ogle_lc_all} indicates that the system becomes redder when brighter.
This colour modulation can be attributed to the presence of a decretion disk around the Be star, as a face-on configuration of the disk is expected to cause this modulation \citep[for similar examples see ][]{2011MNRAS.413.1600R}. 

\section{Discussion}
 \label{discussion}

By monitoring the BeXRB system IGR\,J05007-7047 we were able to promptly trigger an \xmm ToO.  
Timing analysis of the photon events collected during the \xmm ToO reveals coherent X-ray pulsations (see Fig. \ref{fig:psd}) with a period of 38.551(3) s (1$\sigma$), making \lxp the 17$^{th}$ known HMXB pulsar in the LMC \citep[see][ and references therein]{valia15}. 
During the outburst, the X-ray spectrum is well represented by an absorbed power law ($\Gamma=0.85$), 
however we argue for the presence of additional components that can both improve the residuals of the fit and account for known physical processes resulting in X-ray emission from BeXRB systems.
Our analysis indicates that a model composed of an absorbed power-law plus a high-temperature black-body (kT $\sim$ 2 keV) model (Fig. \ref{fig:spec1}) improves the fit. A model composed by an absorbed power law ($\Gamma=0.6$) and a multi-temperature low-temperature disk black body (kT$\sim0.1$ keV) provides an acceptable fit but only improves the residuals at lower energies. 
The tested models can help us to study different emission mechanisms and understand the physical
properties of different regions in the binary system, like the accretion column towards the magnetic poles of the neutron star, or the
inner edge of the accretion disk at the outer parts of the magnetosphere.

The normalization of the best fit model with a high-energy black-body component yields an accretion region with a radius of 0.7 km, and can be interpreted as emission from a polar cap. 
The normalization of the multi-temperature black-body component translates to an emitting region with a size of $\sim200~\rm km$, which could be the inner region of an accretion disk around the NS.
Another possible explanation is that this low temperature multi-temperature black-body component originates from reprocessing of X-rays in the inner part of the accretion disk. Hard X-rays originating from the NS can illuminate an area at distance $R_{ir}$ under a solid angle $\Omega$. This radius is different than the one computed from the normalization of the black-body and can be calculated from $R_{ir}=(4\pi{\sigma}T_{BB}^4/{L_x})^{-1/2}$ \citep{2004ApJ...614..881H}, where $L_x$ is the total X-ray luminosity and $T_{BB}$  is the temperature of the black-body. From fitted parameters we calculate a radius of $\sim600~\rm km$.

Expressing the X-ray luminosity in terms of mass accretion rate we find \hbox{$\dot{M}=3.5\times10^{-10}$~\msun$\rm~year^{-1}$} using a typical accretion efficiency value of 0.2 \citep{2000AstL...26..699S}.
Assuming the magnetospheric radius of the NS has the same size as the accretion disk inner radius we can make an order of magnitude estimation of the magnetic field strength of the NS. Following \citet{2002apa..book.....F}, for an X-ray luminosity of 4\ergs{36}, a magnetospheric radius of 200 km and typical values for the mass and radius ($M_{\rm NS}$=1.4~\msun\ and $R_{\rm NS}$=12.5 km) we derive a magnetic field strength of $10^{10}$~G.   
Moreover we note that the magnetospheric radius is much smaller than the corotation radius of the NS, which is \hbox{$R_{co}=(G~M_{NS}~P^2_{\rm spin}~\pi^{-2}/4)^{1/3}=1.9\times10^4~\rm km $}, where $P_{\rm spin}$ is the spin period of the NS.
If we assume that the accretion originates from a Keplerian disk with inner radius $r$ where the velocity of the disk exceeds the spin angular velocity of the NS, we conclude that the interaction between the Keplerian disk and the NS results in a spin-up of the system (accretor regime).

The 30.77 d optical modulation can be attributed to the orbital period of the NS-Be star system. 
This modulation is associated to the truncation of the Be disk due to gravitational interaction with the NS during an orbital revolution. 
The colour modulation --redder when brighter-- is also a result of changes in the decretion disk around the Be star.
A face-on geometry of the disk is expected to cause the above behaviour \citep{2011MNRAS.413.1600R}. 
Similar behaviour has been encountered in numerous BeXRB systems, like XMMU\,J010743.1-715953 \citep{2012MNRAS.424..282C}, LXP\,168.8 \citep{2013A&A...554A...1M}, LXP\,8.04 \citep{2014A&A...567A.129V} or isolated Be Stars in the SMC \citep{2006A&A...456.1027D}.
The disk adds cooler light to the stellar emission of the B Starwithout obscuring the star itself (face-on geometry), while variations in the disk size or radius within the binary orbit result in a variable SED. 
This is also noticeable in the GROND observations where the optical part of the SED (dominated by the B star) is much less variable than the IR (larger contribution from the Be disk).  

The long term optical variability (super-orbital) is believed to be related to the formation and depletion of the circumstellar disk around the Be star \citep{2011MNRAS.413.1600R}. 
The $\sim30.77~\rm d$ orbital modulation was still present in all the available data, and the same colour modulation --redder when brighter-- was detected (see Fig. \ref{fig:ogle_lc_all}). 
During the OGLE-III phase the system exhibited high long-term variability evident of a variable Be disk. There is no evidence for significant changes in the size of the disk during the OGLE-IV epoch, where the amplitude of orbital modulation in the optical band is comparable to the long term variability of the system.  
Multi-epoch observations in the X-ray and optical would allow for a comparison between the long term X-ray and optical variability, 
unfortunately all the available X-ray observations where performed when the OGLE I band magnitude of the system was between 14.55 and 14.7.

By comparing the available X-ray observations that were performed on times separated by more than one orbital period, but correspond to the same orbital phase we can advocate about the long term X-ray variability of the system.
The \xmm slew detection was performed at a phase when only \swift observations were available (phase 0.87 of Fig. \ref{fig:multi_lc_all} \& \ref{fig:ogle_lc2}). By comparing their fluxes we derive that at the time of the \xmm slew detection the luminosity of the system was higher by a factor of $3.3\pm1.0$ compared to the one measured with \swift.
The \cxo observation of the system was performed on MJD$\sim$53537.1 d corresponding to phase 0.61. The luminosity of the system was found to be $\sim$2.6\ergs{36} in the 0.3-10.0 keV band.
By comparing the reported luminosity to the \swift luminosity corresponding to the same orbital phase we found that the system was brighter by a factor of $\sim1.1$ during the \swift/XRT monitoring performed on November 2014. 
Our data confirm the findings of \citet{2011A&A...529A..30D} that the systems long term X-ray emission is persistent within a variability factor of $\sim$3-4, which is lower than the orbital variability (variability factor $\sim10$). 

Our \swift/XRT monitoring was performed when the average OGLE I band magnitude of the system was $\sim$14.65. Based on the SED modelling we found that in the I band the system has an excess of $\sim$24\% compared to the template flux of a B star with same temperature (see Tab. \ref{tab:sed}). Interestingly this translates to a difference of $\sim$0.23 in magnitude, 
which is exactly the difference between the average I magnitude during our monitoring and the lowest flux level observed in the OGLE I band; I mag of 14.88 at  MJD 52930 d.   
This might be evident of an epoch where the Be disk was almost completely depleted.

\citet{2011A&A...529A..30D} reported the system as a possible wind fed X-ray system, but our results reveal the presence of a cooler component in the SED of the companion star suggesting that this is a NS pulsar with a Be companion. Moreover, the hot thermal excess in the X-ray spectrum adds to the population of BeXRB systems sharing this feature. 
Recently \citet{2013MmSAI..84..626L} reported a population of long spin ($>$200 s), low luminosity (L$\sim$\oergs{34}), X-ray persistent Galactic BeXRBs systems showing this characteristic feature.
A similar hot thermal excess was also found in the X-ray spectra of other BeXRBs in the Magellanic Clouds during moderate outbursts with L$\sim$\oergs{36} \citep{2013A&A...558A..74V,2013MNRAS.436.2054B}.
\lxp is likely another example of such a system, showing evidence of a hot black-body component at higher X-ray luminosity, suggesting the black-body component is more common and not limited to the population originally proposed by \citet{2013MmSAI..84..626L}.
It is not clear if this feature is model dependent, but the physical parameters derived from such a phenomenological spectral treatment where the degeneracy of the fitted parameters is significant should be treated as order of magnitude approximations. 
Yet the study of such systems in the Magellanic Clouds has the major advantage of the low column density in the line of sight, thus through the study of these systems we are able to improve the modelling of their X-ray spectrum and derive stringent constraints on our findings.

\begin{figure}
    \resizebox{\hsize}{!}{\includegraphics[angle=0,clip=]{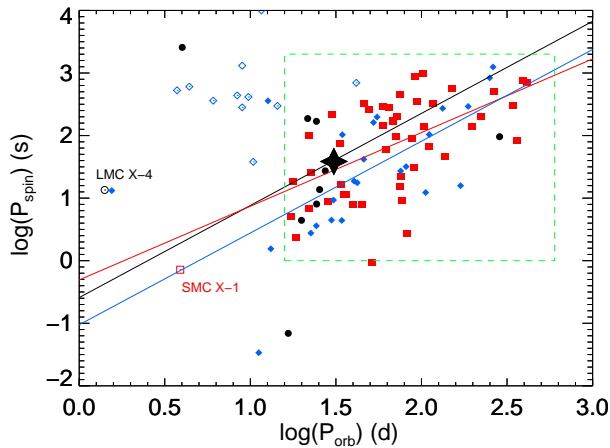}}
    \vspace{-0.5cm}
  \caption{Corbet diagram of the LMC (black circles), SMC (red squares), and Milky Way (blue diamonds) HMXB pulsars. Open symbols denote SG-XRB systems while filled indicate BeXRB systems.
  The black star points to the position of \lxp.
  The green box indicates the locus where most of the BeXRB pulsars are found. The best fit lines of the three groups are produced from points within this area and have slopes of $\sim$1.5, $\sim$1.2 and $\sim$1.5 respectively.
   }
  \label{fig:corbet}
\end{figure}

The position of \lxp in the Corbet diagram \citep{2011Natur.479..372K} lies close to the best fit line derived from the population of confirmed MC pulsars  (see Fig. \ref{fig:corbet}). 
Looking at the properties of other BeXRB systems located near \lxp in the $P_{\rm spin}$ vs. $P_{\rm orbit}$ diagram, we found no notable similarities between them and the newly reported system. Most of these systems show much larger long term X-ray variability, with variability factors $\sim$1000, as expected for such short-spin systems \citep{2016A&A...586A..81H}.
However since \lxp is a newly discovered system, future observations might reveal a higher variability factor.
Moreover the lack of deep X-ray observations during their peak luminosity prohibits us from comparing their X-ray spectral properties. 
SXP\,25.55 \citep{2008A&A...485..177H} with an orbital period of 22.5 d \citep{2011MNRAS.413.1600R} is an SMC BeXRB pulsar located very close to the position of \lxp in the Corbet diagram. An \xmm spectrum is available for this system, but its peak luminosity is about 30 times smaller than that of \lxp, while its spectrum is affected by stronger absorption.  
LXP\,28.28 \citep{2016MNRAS.456..845S} is the closest LMC system having an orbital period of $\sim27.1$ d. \citep{2015MNRAS.447.1630C}. 

\section{Conclusion}
 \label{conclusion}
A triggered \xmm ToO observation of the candidate BeXRB system \igr located in the LMC revealed coherent X-ray pulsations with a period of 38.551(3) s. The X-ray spectrum of the system can be described by an absorbed power-law ($\Gamma\sim0.63$) plus a hot black body component ($\sim2.04$ keV) that can be interpreted as black body emission from a small region, most likely the polar cap of the NS. Additionally there is a statistically significant evidence for a soft excess that can originate from a region close to the magnetospheric radius and is a result of reprocessing of hard X-rays emitted by the accretion column of the NS. Analysis of optical data confirms the 30.776(5) d orbital period that was previously reported. By using simultaneously obtained optical and NIR data we constructed the SED of the binary system that reveals the presence of a decretion disk around the massive optical companion of the NS.   
Our analysis confirms \lxp as the 17$^{th}$ HMXB pulsar in this star forming galaxy.

\section*{Acknowledgments}
We would like to thank the anonymous referee for his/her critical review and suggestions that helped improve the manuscript.  
The \xmm\ project is supported by the Bundesministerium f\"ur Wirtschaft und
Technologie\,/\,Deutsches Zentrum f\"ur Luft- und Raumfahrt (BMWi/DLR, FKZ 50 OX
0001) and the Max-Planck Society. 
The OGLE project has received funding from the National Science Centre,
Poland, grant MAESTRO 2014/14/A/ST9/00121 to AU.
We thank the \swift team for accepting and carefully scheduling the target of opportunity observations, and we acknowledge the use of public data from the \swift data archive. 
G.\,V. acknowledge support from the BMWi/DLR grant FKZ 50 OR 1208.

\bibliographystyle{mnras}
\bibliography{IGR_J05007-7047}

\begin{thebibliography}{}
\makeatletter
\relax
\def\mn@urlcharsother{\let\do\@makeother \do\$\do\&\do\#\do\^\do\_\do\%\do\~}
\def\mn@doi{\begingroup\mn@urlcharsother \@ifnextchar [ {\mn@doi@}
  {\mn@doi@[]}}
\def\mn@doi@[#1]#2{\def\@tempa{#1}\ifx\@tempa\@empty \href
  {http://dx.doi.org/#2} {doi:#2}\else \href {http://dx.doi.org/#2} {#1}\fi
  \endgroup}
\def\mn@eprint#1#2{\mn@eprint@#1:#2::\@nil}
\def\mn@eprint@arXiv#1{\href {http://arxiv.org/abs/#1} {{\tt arXiv:#1}}}
\def\mn@eprint@dblp#1{\href {http://dblp.uni-trier.de/rec/bibtex/#1.xml}
  {dblp:#1}}
\def\mn@eprint@#1:#2:#3:#4\@nil{\def\@tempa {#1}\def\@tempb {#2}\def\@tempc
  {#3}\ifx \@tempc \@empty \let \@tempc \@tempb \let \@tempb \@tempa \fi \ifx
  \@tempb \@empty \def\@tempb {arXiv}\fi \@ifundefined
  {mn@eprint@\@tempb}{\@tempb:\@tempc}{\expandafter \expandafter \csname
  mn@eprint@\@tempb\endcsname \expandafter{\@tempc}}}

\bibitem[\protect\citeauthoryear{{Antoniou} \& {Zezas}}{{Antoniou} \&
  {Zezas}}{2016}]{valia15}
{Antoniou} V.,  {Zezas} A.,  2016, preprint, \href
  {http://adsabs.harvard.edu/abs/2016arXiv160308011A} {} (\mn@eprint {arXiv}
  {1603.08011})

\bibitem[\protect\citeauthoryear{{Arnaud}}{{Arnaud}}{1996}]{1996ASPC..101...17%
A}
{Arnaud} K.~A.,  1996, in {Jacoby} G.~H.,  {Barnes} J.,  eds,  Astronomical
  Society of the Pacific Conference Series Vol. 101, Astronomical Data Analysis
  Software and Systems V. p.~17

\bibitem[\protect\citeauthoryear{{Barthelmy} et~al.,}{{Barthelmy}
  et~al.}{2005}]{2005SSRv..120..143B}
{Barthelmy} S.~D.,  et~al., 2005, \mn@doi [\ssr] {10.1007/s11214-005-5096-3},
  \href {http://adsabs.harvard.edu/abs/2005SSRv..120..143B} {120, 143}

\bibitem[\protect\citeauthoryear{{Bartlett}, {Coe}  \& {Ho}}{{Bartlett}
  et~al.}{2013}]{2013MNRAS.436.2054B}
{Bartlett} E.~S.,  {Coe} M.~J.,   {Ho} W.~C.~G.,  2013, \mn@doi [\mnras]
  {10.1093/mnras/stt1711}, \href
  {http://adsabs.harvard.edu/abs/2013MNRAS.436.2054B} {436, 2054}

\bibitem[\protect\citeauthoryear{{Castelli}, {Gratton}  \& {Kurucz}}{{Castelli}
  et~al.}{1997}]{1997A&A...318..841C}
{Castelli} F.,  {Gratton} R.~G.,   {Kurucz} R.~L.,  1997, \aap, \href
  {http://adsabs.harvard.edu/abs/1997A%26A...318..841C} {318, 841}

\bibitem[\protect\citeauthoryear{{Coe}, {Edge}, {Galache}  \& {McBride}}{{Coe}
  et~al.}{2005}]{2005MNRAS.356..502C}
{Coe} M.~J.,  {Edge} W.~R.~T.,  {Galache} J.~L.,   {McBride} V.~A.,  2005,
  \mn@doi [\mnras] {10.1111/j.1365-2966.2004.08467.x}, \href
  {http://adsabs.harvard.edu/abs/2005MNRAS.356..502C} {356, 502}

\bibitem[\protect\citeauthoryear{{Coe} et~al.,}{{Coe}
  et~al.}{2012}]{2012MNRAS.424..282C}
{Coe} M.~J.,  et~al., 2012, \mn@doi [\mnras]
  {10.1111/j.1365-2966.2012.21193.x}, \href
  {http://adsabs.harvard.edu/abs/2012MNRAS.424..282C} {424, 282}

\bibitem[\protect\citeauthoryear{{Coe}, {Finger}, {Bartlett}  \&
  {Udalski}}{{Coe} et~al.}{2015}]{2015MNRAS.447.1630C}
{Coe} M.~J.,  {Finger} M.,  {Bartlett} E.~S.,   {Udalski} A.,  2015, \mn@doi
  [\mnras] {10.1093/mnras/stu2562}, \href
  {http://adsabs.harvard.edu/abs/2015MNRAS.447.1630C} {447, 1630}

\bibitem[\protect\citeauthoryear{{Cox}}{{Cox}}{2000}]{2000asqu.book.....C}
{Cox} A.~N.,  2000, {Allen's astrophysical quantities}

\bibitem[\protect\citeauthoryear{{D'A{\`i}}, {La Parola}, {Cusumano},
  {Segreto}, {Romano}, {Vercellone}  \& {Robba}}{{D'A{\`i}}
  et~al.}{2011}]{2011A&A...529A..30D}
{D'A{\`i}} A.,  {La Parola} V.,  {Cusumano} G.,  {Segreto} A.,  {Romano} P.,
  {Vercellone} S.,   {Robba} N.~R.,  2011, \mn@doi [\aap]
  {10.1051/0004-6361/201016401}, \href
  {http://cdsads.u-strasbg.fr/abs/2011A%26A...529A..30D} {529, A30}

\bibitem[\protect\citeauthoryear{{Davies}}{{Davies}}{1990}]{1990MNRAS.244...93%
D}
{Davies} S.~R.,  1990, \mnras, \href
  {http://adsabs.harvard.edu/abs/1990MNRAS.244...93D} {244, 93}

\bibitem[\protect\citeauthoryear{{Dickey} \& {Lockman}}{{Dickey} \&
  {Lockman}}{1990}]{1990ARA&A..28..215D}
{Dickey} J.~M.,  {Lockman} F.~J.,  1990, \mn@doi [\araa]
  {10.1146/annurev.aa.28.090190.001243}, \href
  {http://adsabs.harvard.edu/abs/1990ARA%26A..28..215D} {28, 215}

\bibitem[\protect\citeauthoryear{{Dougherty}, {Waters}, {Burki}, {Cote},
  {Cramer}, {van Kerkwijk}  \& {Taylor}}{{Dougherty}
  et~al.}{1994}]{1994A&A...290..609D}
{Dougherty} S.~M.,  {Waters} L.~B.~F.~M.,  {Burki} G.,  {Cote} J.,  {Cramer}
  N.,  {van Kerkwijk} M.~H.,   {Taylor} A.~R.,  1994, \aap, \href
  {http://adsabs.harvard.edu/abs/1994A%26A...290..609D} {290}

\bibitem[\protect\citeauthoryear{{Frank}, {King}  \& {Raine}}{{Frank}
  et~al.}{2002}]{2002apa..book.....F}
{Frank} J.,  {King} A.,   {Raine} D.~J.,  2002, {Accretion Power in
  Astrophysics: Third Edition}

\bibitem[\protect\citeauthoryear{{Gehrels} et~al.,}{{Gehrels}
  et~al.}{2004}]{2004ApJ...611.1005G}
{Gehrels} N.,  et~al., 2004, \mn@doi [\apj] {10.1086/422091}, \href
  {http://adsabs.harvard.edu/abs/2004ApJ...611.1005G} {611, 1005}

\bibitem[\protect\citeauthoryear{{Graczyk} et~al.,}{{Graczyk}
  et~al.}{2014}]{2014ApJ...780...59G}
{Graczyk} D.,  et~al., 2014, \mn@doi [\apj] {10.1088/0004-637X/780/1/59}, \href
  {http://adsabs.harvard.edu/abs/2014ApJ...780...59G} {780, 59}

\bibitem[\protect\citeauthoryear{{Gregory} \& {Loredo}}{{Gregory} \&
  {Loredo}}{1996}]{1996ApJ...473.1059G}
{Gregory} P.~C.,  {Loredo} T.~J.,  1996, \mn@doi [\apj] {10.1086/178215}, \href
  {http://adsabs.harvard.edu/abs/1996ApJ...473.1059G} {473, 1059}

\bibitem[\protect\citeauthoryear{{Greiner} et~al.,}{{Greiner}
  et~al.}{2008}]{2008PASP..120..405G}
{Greiner} J.,  et~al., 2008, \mn@doi [\pasp] {10.1086/587032}, \href
  {http://adsabs.harvard.edu/abs/2008PASP..120..405G} {120, 405}

\bibitem[\protect\citeauthoryear{{Haberl} \& {Sturm}}{{Haberl} \&
  {Sturm}}{2016}]{2016A&A...586A..81H}
{Haberl} F.,  {Sturm} R.,  2016, \mn@doi [\aap] {10.1051/0004-6361/201527326},
  \href {http://adsabs.harvard.edu/abs/2016A%26A...586A..81H} {586, A81}

\bibitem[\protect\citeauthoryear{{Haberl}, {Eger}, {Pietsch}, {Corbet}  \&
  {Sasaki}}{{Haberl} et~al.}{2008a}]{2008A&A...485..177H}
{Haberl} F.,  {Eger} P.,  {Pietsch} W.,  {Corbet} R.~H.~D.,   {Sasaki} M.,
  2008a, \mn@doi [\aap] {10.1051/0004-6361:200809540}, \href
  {http://adsabs.harvard.edu/abs/2008A%26A...485..177H} {485, 177}

\bibitem[\protect\citeauthoryear{{Haberl}, {Eger}  \& {Pietsch}}{{Haberl}
  et~al.}{2008b}]{2008A&A...489..327H}
{Haberl} F.,  {Eger} P.,   {Pietsch} W.,  2008b, \mn@doi [\aap]
  {10.1051/0004-6361:200810100}, \href
  {http://adsabs.harvard.edu/abs/2008A%26A...489..327H} {489, 327}

\bibitem[\protect\citeauthoryear{{Haberl} et~al.,}{{Haberl}
  et~al.}{2012}]{2012A&A...545A.128H}
{Haberl} F.,  et~al., 2012, \mn@doi [\aap] {10.1051/0004-6361/201219758}, \href
  {http://adsabs.harvard.edu/abs/2012A%26A...545A.128H} {545, A128}

\bibitem[\protect\citeauthoryear{{Hagen}, {Siegel}, {Gronwall}, {Hoversten}  \&
  {Immler}}{{Hagen} et~al.}{2014}]{2014AAS...22344230H}
{Hagen} L.,  {Siegel} M.,  {Gronwall} C.,  {Hoversten} E.~A.,   {Immler} S.,
  2014, in American Astronomical Society Meeting Abstracts \#223. p. \#442.30

\bibitem[\protect\citeauthoryear{{Hickox}, {Narayan}  \& {Kallman}}{{Hickox}
  et~al.}{2004}]{2004ApJ...614..881H}
{Hickox} R.~C.,  {Narayan} R.,   {Kallman} T.~R.,  2004, \mn@doi [\apj]
  {10.1086/423928}, \href {http://adsabs.harvard.edu/abs/2004ApJ...614..881H}
  {614, 881}

\bibitem[\protect\citeauthoryear{{Horne} \& {Baliunas}}{{Horne} \&
  {Baliunas}}{1986}]{1986ApJ...302..757H}
{Horne} J.~H.,  {Baliunas} S.~L.,  1986, \mn@doi [\apj] {10.1086/164037}, \href
  {http://adsabs.harvard.edu/abs/1986ApJ...302..757H} {302, 757}

\bibitem[\protect\citeauthoryear{{Kalberla}, {Burton}, {Hartmann}, {Arnal},
  {Bajaja}, {Morras}  \& {P{\"o}ppel}}{{Kalberla}
  et~al.}{2005}]{2005A&A...440..775K}
{Kalberla} P.~M.~W.,  {Burton} W.~B.,  {Hartmann} D.,  {Arnal} E.~M.,  {Bajaja}
  E.,  {Morras} R.,   {P{\"o}ppel} W.~G.~L.,  2005, \mn@doi [\aap]
  {10.1051/0004-6361:20041864}, \href
  {http://adsabs.harvard.edu/abs/2005A%26A...440..775K} {440, 775}

\bibitem[\protect\citeauthoryear{{Knigge}, {Coe}  \& {Podsiadlowski}}{{Knigge}
  et~al.}{2011}]{2011Natur.479..372K}
{Knigge} C.,  {Coe} M.~J.,   {Podsiadlowski} P.,  2011, \mn@doi [\nat]
  {10.1038/nature10529}, \href
  {http://adsabs.harvard.edu/abs/2011Natur.479..372K} {479, 372}

\bibitem[\protect\citeauthoryear{{Kr{\"u}hler} et~al.,}{{Kr{\"u}hler}
  et~al.}{2008}]{2008ApJ...685..376K}
{Kr{\"u}hler} T.,  et~al., 2008, \mn@doi [\apj] {10.1086/590240}, \href
  {http://adsabs.harvard.edu/abs/2008ApJ...685..376K} {685, 376}

\bibitem[\protect\citeauthoryear{{Kurucz}}{{Kurucz}}{1979}]{1979ApJS...40....1%
K}
{Kurucz} R.~L.,  1979, \mn@doi [\apjs] {10.1086/190589}, \href
  {http://cdsads.u-strasbg.fr/abs/1979ApJS...40....1K} {40, 1}

\bibitem[\protect\citeauthoryear{{La Palombara}, {Mereghetti}, {Sidoli},
  {Tiengo}  \& {Esposito}}{{La Palombara} et~al.}{2013a}]{2013arXiv1301.5120L}
{La Palombara} N.,  {Mereghetti} S.,  {Sidoli} L.,  {Tiengo} A.,   {Esposito}
  P.,  2013a, preprint, \href
  {http://adsabs.harvard.edu/abs/2013arXiv1301.5120L} {} (\mn@eprint {arXiv}
  {1301.5120})

\bibitem[\protect\citeauthoryear{{La Palombara}, {Mereghetti}, {Sidoli},
  {Tiengo}  \& {Esposito}}{{La Palombara} et~al.}{2013b}]{2013MmSAI..84..626L}
{La Palombara} N.,  {Mereghetti} S.,  {Sidoli} L.,  {Tiengo} A.,   {Esposito}
  P.,  2013b, \memsai, \href
  {http://adsabs.harvard.edu/abs/2013MmSAI..84..626L} {84, 626}

\bibitem[\protect\citeauthoryear{{Larsson}}{{Larsson}}{1996}]{1996A&AS..117..1%
97L}
{Larsson} S.,  1996, \aaps, \href
  {http://adsabs.harvard.edu/abs/1996A%26AS..117..197L} {117, 197}

\bibitem[\protect\citeauthoryear{{Leahy}, {Darbro}, {Elsner}, {Weisskopf},
  {Kahn}, {Sutherland}  \& {Grindlay}}{{Leahy}
  et~al.}{1983}]{1983ApJ...266..160L}
{Leahy} D.~A.,  {Darbro} W.,  {Elsner} R.~F.,  {Weisskopf} M.~C.,  {Kahn} S.,
  {Sutherland} P.~G.,   {Grindlay} J.~E.,  1983, \mn@doi [\apj]
  {10.1086/160766}, \href {http://adsabs.harvard.edu/abs/1983ApJ...266..160L}
  {266, 160}

\bibitem[\protect\citeauthoryear{{Maggi}, {Haberl}, {Sturm}, {Pietsch}, {Rau},
  {Greiner}, {Udalski}  \& {Sasaki}}{{Maggi}
  et~al.}{2013}]{2013A&A...554A...1M}
{Maggi} P.,  {Haberl} F.,  {Sturm} R.,  {Pietsch} W.,  {Rau} A.,  {Greiner} J.,
   {Udalski} A.,   {Sasaki} M.,  2013, \mn@doi [\aap]
  {10.1051/0004-6361/201321238}, \href
  {http://adsabs.harvard.edu/abs/2013A%26A...554A...1M} {554, A1}

\bibitem[\protect\citeauthoryear{{Masetti} et~al.,}{{Masetti}
  et~al.}{2006}]{2006A&A...459...21M}
{Masetti} N.,  et~al., 2006, \mn@doi [\aap] {10.1051/0004-6361:20066055}, \href
  {http://adsabs.harvard.edu/abs/2006A%26A...459...21M} {459, 21}

\bibitem[\protect\citeauthoryear{{Misselt}, {Clayton}  \& {Gordon}}{{Misselt}
  et~al.}{1999}]{1999ApJ...515..128M}
{Misselt} K.~A.,  {Clayton} G.~C.,   {Gordon} K.~D.,  1999, \mn@doi [\apj]
  {10.1086/307010}, \href {http://adsabs.harvard.edu/abs/1999ApJ...515..128M}
  {515, 128}

\bibitem[\protect\citeauthoryear{{Monet} et~al.,}{{Monet}
  et~al.}{2003}]{2003AJ....125..984M}
{Monet} D.~G.,  et~al., 2003, \mn@doi [\aj] {10.1086/345888}, \href
  {http://adsabs.harvard.edu/abs/2003AJ....125..984M} {125, 984}

\bibitem[\protect\citeauthoryear{{Pietrzy{\'n}ski} et~al.,}{{Pietrzy{\'n}ski}
  et~al.}{2013}]{2013Natur.495...76P}
{Pietrzy{\'n}ski} G.,  et~al., 2013, \mn@doi [\nat] {10.1038/nature11878},
  \href {http://adsabs.harvard.edu/abs/2013Natur.495...76P} {495, 76}

\bibitem[\protect\citeauthoryear{{Rajoelimanana}, {Charles}  \&
  {Udalski}}{{Rajoelimanana} et~al.}{2011}]{2011MNRAS.413.1600R}
{Rajoelimanana} A.~F.,  {Charles} P.~A.,   {Udalski} A.,  2011, \mn@doi
  [\mnras] {10.1111/j.1365-2966.2011.18243.x}, \href
  {http://adsabs.harvard.edu/abs/2011MNRAS.413.1600R} {413, 1600}

\bibitem[\protect\citeauthoryear{{Reig}}{{Reig}}{2011}]{2011Ap&SS.332....1R}
{Reig} P.,  2011, \mn@doi [\apss] {10.1007/s10509-010-0575-8}, \href
  {http://adsabs.harvard.edu/abs/2011Ap%26SS.332....1R} {332, 1}

\bibitem[\protect\citeauthoryear{{Rolleston}, {Trundle}  \&
  {Dufton}}{{Rolleston} et~al.}{2002}]{2002A&A...396...53R}
{Rolleston} W.~R.~J.,  {Trundle} C.,   {Dufton} P.~L.,  2002, \mn@doi [\aap]
  {10.1051/0004-6361:20021088}, \href
  {http://adsabs.harvard.edu/abs/2002A%26A...396...53R} {396, 53}

\bibitem[\protect\citeauthoryear{{Saxton}, {Read}, {Esquej}, {Freyberg},
  {Altieri}  \& {Bermejo}}{{Saxton} et~al.}{2008}]{2008A&A...480..611S}
{Saxton} R.~D.,  {Read} A.~M.,  {Esquej} P.,  {Freyberg} M.~J.,  {Altieri} B.,
   {Bermejo} D.,  2008, \mn@doi [\aap] {10.1051/0004-6361:20079193}, \href
  {http://adsabs.harvard.edu/abs/2008A%26A...480..611S} {480, 611}

\bibitem[\protect\citeauthoryear{{Sazonov}, {Churazov}, {Revnivtsev},
  {Vikhlinin}  \& {Sunyaev}}{{Sazonov} et~al.}{2005}]{2005A&A...444L..37S}
{Sazonov} S.,  {Churazov} E.,  {Revnivtsev} M.,  {Vikhlinin} A.,   {Sunyaev}
  R.,  2005, \mn@doi [\aap] {10.1051/0004-6361:200500205}, \href
  {http://adsabs.harvard.edu/abs/2005A%26A...444L..37S} {444, L37}

\bibitem[\protect\citeauthoryear{{Scargle}}{{Scargle}}{1982}]{1982ApJ...263..8%
35S}
{Scargle} J.~D.,  1982, \mn@doi [\apj] {10.1086/160554}, \href
  {http://adsabs.harvard.edu/abs/1982ApJ...263..835S} {263, 835}

\bibitem[\protect\citeauthoryear{{Schlafly} \& {Finkbeiner}}{{Schlafly} \&
  {Finkbeiner}}{2011}]{2011ApJ...737..103S}
{Schlafly} E.~F.,  {Finkbeiner} D.~P.,  2011, \mn@doi [\apj]
  {10.1088/0004-637X/737/2/103}, \href
  {http://adsabs.harvard.edu/abs/2011ApJ...737..103S} {737, 103}

\bibitem[\protect\citeauthoryear{{Schulz} \& {Mudelsee}}{{Schulz} \&
  {Mudelsee}}{2002}]{2002CG.....28..421S}
{Schulz} M.,  {Mudelsee} M.,  2002, \mn@doi [Computers and Geosciences]
  {10.1016/S0098-3004(01)00044-9}, \href
  {http://adsabs.harvard.edu/abs/2002CG.....28..421S} {28, 421}

\bibitem[\protect\citeauthoryear{{Sibgatullin} \& {Sunyaev}}{{Sibgatullin} \&
  {Sunyaev}}{2000}]{2000AstL...26..699S}
{Sibgatullin} N.~R.,  {Sunyaev} R.~A.,  2000, \mn@doi [Astronomy Letters]
  {10.1134/1.1323277}, \href
  {http://adsabs.harvard.edu/abs/2000AstL...26..699S} {26, 699}

\bibitem[\protect\citeauthoryear{{Skrutskie} et~al.,}{{Skrutskie}
  et~al.}{2006}]{2006AJ....131.1163S}
{Skrutskie} M.~F.,  et~al., 2006, \mn@doi [\aj] {10.1086/498708}, \href
  {http://adsabs.harvard.edu/abs/2006AJ....131.1163S} {131, 1163}

\bibitem[\protect\citeauthoryear{{Stuhlinger} et~al.,}{{Stuhlinger}
  et~al.}{2006}]{2006ESASP.604..937S}
{Stuhlinger} M.,  et~al., 2006, in {Wilson} A.,  ed.,  ESA Special Publication
  Vol. 604, The X-ray Universe 2005. p.~937 (\mn@eprint {}
  {arXiv:astro-ph/0511395})

\bibitem[\protect\citeauthoryear{{Sturm} et~al.,}{{Sturm}
  et~al.}{2013}]{2013A&A...558A...3S}
{Sturm} R.,  et~al., 2013, \mn@doi [\aap] {10.1051/0004-6361/201219935}, \href
  {http://adsabs.harvard.edu/abs/2013A%26A...558A...3S} {558, A3}

\bibitem[\protect\citeauthoryear{{Ubertini} et~al.,}{{Ubertini}
  et~al.}{2003}]{2003A&A...411L.131U}
{Ubertini} P.,  et~al., 2003, \mn@doi [\aap] {10.1051/0004-6361:20031224},
  \href {http://adsabs.harvard.edu/abs/2003A%26A...411L.131U} {411, L131}

\bibitem[\protect\citeauthoryear{{Udalski}}{{Udalski}}{2008}]{2008AcA....58..1%
87U}
{Udalski} A.,  2008, \actaa, \href
  {http://adsabs.harvard.edu/abs/2008AcA....58..187U} {58, 187}

\bibitem[\protect\citeauthoryear{{Udalski}, {Szymanski}, {Kaluzny}, {Kubiak}
  \& {Mateo}}{{Udalski} et~al.}{1992}]{1992AcA....42..253U}
{Udalski} A.,  {Szymanski} M.,  {Kaluzny} J.,  {Kubiak} M.,   {Mateo} M.,
  1992, \actaa, \href {http://adsabs.harvard.edu/abs/1992AcA....42..253U} {42,
  253}

\bibitem[\protect\citeauthoryear{{Udalski}, {Szyma{\'n}ski}  \&
  {Szyma{\'n}ski}}{{Udalski} et~al.}{2015}]{2015AcA....65....1U}
{Udalski} A.,  {Szyma{\'n}ski} M.~K.,   {Szyma{\'n}ski} G.,  2015, \actaa,
  \href {http://adsabs.harvard.edu/abs/2015AcA....65....1U} {65, 1}

\bibitem[\protect\citeauthoryear{{Vasilopoulos}, {Maggi}, {Haberl}, {Sturm},
  {Pietsch}, {Bartlett}  \& {Coe}}{{Vasilopoulos}
  et~al.}{2013}]{2013A&A...558A..74V}
{Vasilopoulos} G.,  {Maggi} P.,  {Haberl} F.,  {Sturm} R.,  {Pietsch} W.,
  {Bartlett} E.~S.,   {Coe} M.~J.,  2013, \mn@doi [\aap]
  {10.1051/0004-6361/201322335}, \href
  {http://adsabs.harvard.edu/abs/2013A%26A...558A..74V} {558, A74}

\bibitem[\protect\citeauthoryear{{Vasilopoulos}, {Haberl}, {Sturm}, {Maggi}  \&
  {Udalski}}{{Vasilopoulos} et~al.}{2014}]{2014A&A...567A.129V}
{Vasilopoulos} G.,  {Haberl} F.,  {Sturm} R.,  {Maggi} P.,   {Udalski} A.,
  2014, \mn@doi [\aap] {10.1051/0004-6361/201423934}, \href
  {http://adsabs.harvard.edu/abs/2014A%26A...567A.129V} {567, A129}

\bibitem[\protect\citeauthoryear{{Verner}, {Ferland}, {Korista}  \&
  {Yakovlev}}{{Verner} et~al.}{1996}]{1996ApJ...465..487V}
{Verner} D.~A.,  {Ferland} G.~J.,  {Korista} K.~T.,   {Yakovlev} D.~G.,  1996,
  \mn@doi [\apj] {10.1086/177435}, \href
  {http://adsabs.harvard.edu/abs/1996ApJ...465..487V} {465, 487}

\bibitem[\protect\citeauthoryear{{Watson} et~al.,}{{Watson}
  et~al.}{2009}]{2009A&A...493..339W}
{Watson} M.~G.,  et~al., 2009, \mn@doi [\aap] {10.1051/0004-6361:200810534},
  \href {http://adsabs.harvard.edu/abs/2009A%26A...493..339W} {493, 339}

\bibitem[\protect\citeauthoryear{{Wilms}, {Allen}  \& {McCray}}{{Wilms}
  et~al.}{2000}]{2000ApJ...542..914W}
{Wilms} J.,  {Allen} A.,   {McCray} R.,  2000, \mn@doi [\apj] {10.1086/317016},
  \href {http://adsabs.harvard.edu/abs/2000ApJ...542..914W} {542, 914}

\bibitem[\protect\citeauthoryear{{Winkler} et~al.,}{{Winkler}
  et~al.}{2003}]{2003A&A...411L...1W}
{Winkler} C.,  et~al., 2003, \mn@doi [\aap] {10.1051/0004-6361:20031288}, \href
  {http://adsabs.harvard.edu/abs/2003A%26A...411L...1W} {411, L1}

\bibitem[\protect\citeauthoryear{{Yolda{\c s}}, {Kr{\"u}hler}, {Greiner},
  {Yolda{\c s}}, {Clemens}, {Szokoly}, {Primak}  \& {Klose}}{{Yolda{\c s}}
  et~al.}{2008}]{2008AIPC.1000..227Y}
{Yolda{\c s}} A.~K.,  {Kr{\"u}hler} T.,  {Greiner} J.,  {Yolda{\c s}} A.,
  {Clemens} C.,  {Szokoly} G.,  {Primak} N.,   {Klose} S.,  2008, in {Galassi}
  M.,  {Palmer} D.,   {Fenimore} E.,  eds,  American Institute of Physics
  Conference Series Vol. 1000, American Institute of Physics Conference Series.
  pp 227--231, \mn@doi{10.1063/1.2943450}

\bibitem[\protect\citeauthoryear{{York} et~al.,}{{York}
  et~al.}{2000}]{2000AJ....120.1579Y}
{York} D.~G.,  et~al., 2000, \mn@doi [\aj] {10.1086/301513}, \href
  {http://adsabs.harvard.edu/abs/2000AJ....120.1579Y} {120, 1579}

\bibitem[\protect\citeauthoryear{{{\c S}ahiner}, {Serim}, {Baykal}  \&
  {{\.I}nam}}{{{\c S}ahiner} et~al.}{2016}]{2016MNRAS.456..845S}
{{\c S}ahiner} {\c S}.,  {Serim} M.~M.,  {Baykal} A.,   {{\.I}nam} S.~{\c C}.,
  2016, \mn@doi [\mnras] {10.1093/mnras/stv2775}, \href
  {http://adsabs.harvard.edu/abs/2016MNRAS.456..845S} {456, 845}

\bibitem[\protect\citeauthoryear{{de Wit}, {Lamers}, {Marquette}  \&
  {Beaulieu}}{{de Wit} et~al.}{2006}]{2006A&A...456.1027D}
{de Wit} W.~J.,  {Lamers} H.~J.~G.~L.~M.,  {Marquette} J.~B.,   {Beaulieu}
  J.~P.,  2006, \mn@doi [\aap] {10.1051/0004-6361:20065137}, \href
  {http://adsabs.harvard.edu/abs/2006A%26A...456.1027D} {456, 1027}

\makeatother
\end{thebibliography}

\end{document}